\documentclass[a4paper,12pt]{article}
\usepackage{epsfig}
\usepackage[dvips,usenames]{color}
\usepackage{graphicx}

\newlength{\dinwidth}
\newlength{\dinmargin}
\newcommand{\spur}[1]{\not\! #1 \,}
\def\pslash{\rlap{\hspace{0.02cm}/}{p}}
\def\kslash{\rlap{\hspace{0.02cm}/}{k}}

\begin{document}

\title{\bf Reexamining charmless $B\to PV$ decays in QCD factorization approach}
\bigskip

\author{ Xinqiang Li$^{1,2,3}$ and Yadong Yang$^{1}$~\footnote{ Corresponding author.
E-mail address: yangyd@henannu.edu.cn}
\\
{ $^1$\small Department of Physics, Henan Normal University,
Xinxiang, Henan 453007, P.~R. China}
\\
{ $^2$\small Institute of Theoretical Physics, Chinese Academy of
Sciences, Beijing, 100080, P.~R. China}
\\
{ $^3$\small Graduate School of the Chinese Academy of Sciences,
Beijing, 100039, P.~R. China} }

\maketitle
\begin{picture}(0,0)
\put(305,290){\sf hep-ph/0602224}
\end{picture}
\bigskip\bigskip
\maketitle \vspace{-1.5cm}
\begin{abstract}
Using the QCD factorization approach, we reexamine the two-body
hadronic charmless $B$-meson decays to final states involving a
pseudoscalar~($P$) and a vector~($V$) meson, with inclusion of the
penguin contractions of spectator-scattering amplitudes induced by
the $b\to D g^\ast g^\ast$~(where $D=d$ or $s$, and $g^\ast$
denotes an off-shell gluon) transitions, which are of order
$\alpha_s^2$. Their impacts on the $CP$-averaged branching ratios
and $CP$-violating asymmetries are  examined. We find that
these higher order penguin contraction contributions have
significant impacts on some specific decay modes.  Since $B\to \pi
K^{\ast}$, $K \rho$ decays involve the same electro-weak physics
as $B\to \pi K$ puzzles, we present a detailed analysis of these
decays and find that the five R-ratios for $B\to \pi K^{\ast}$, $K
\rho$ system are in agreement with experimental data except for
$R(\pi K^*)$. Generally, these new contributions are found to be
important for penguin-dominated $B\to PV$ decays.
\end{abstract}


\newpage
\section{Introduction}

The study of hadronic charmless $B$-meson decays can provide not
only an interesting avenue to understand the $CP$ violation and
the flavor mixing of quark sector in the Standard Model~(SM), but
also powerful means to probe different new physics scenarios
beyond the SM. With the operation of $B$-factory experiments, huge
amount of experimental data on hadronic $B$-meson decays has been
analyzed with appreciative precision. To account for the
experimental data, theorists are urged to gain deep insight into
the mechanism of rare hadronic $B$-meson decays, and to reduce
theoretical uncertainties in determining the flavor parameters of
the SM from experimental measurements.

In the past  years, much progress has been made in understanding
the hadronic charmless $B$-meson decays: several novel methods,
such as the ``naive" factorization~(NF)~\cite{BSW}, the
perturbative QCD method~(PQCD)~\cite{pqcd}, the QCD
factorization~(QCDF)~\cite{bbns1}, and the soft collinear
effective theory~(SCET)~\cite{scet}, have been proposed; in
addition, some model-independent methods based on (approximate)
flavor symmetries have also been used to analyze the rare hadronic
$B$-meson decays~\cite{Wu:2005hi,Chiang:2003pm}. These methods
usually have quite different understandings of the rare hadronic
$B$-meson decays, and hence the corresponding predictions are also
quite different. General comparison between these various methods
can be found, for example, in Ref.~\cite{Keum:2000ms}. Since we
shall adopt the QCDF approach in this paper, we would only focus
on this approach below.

The QCDF approach, put forward by Beneke \textit{et al.} a few
years ago, has been used widely to analyze the two-body hadronic
$B$-meson
decays~\cite{bbns1,bbns2,Muta:2000ti,Yang:2000xn,Du:2002up,bbns3,Li:2005wx}.
The essence of the approach can be summarized as follows: since
the $b$ quark mass is much larger than the characteristic scale of
hadronic interaction, $\Lambda_{\rm QCD}$, to leading power in the
heavy quark expansion, the hadronic matrix elements relevant to
two-body hadronic $B$-meson decays can be factorized into
perturbatively calculable hard scattering kernels and universal
non-perturbative parts parameterized by the form factors and the
meson light cone distribution amplitudes~(LCDAs). This scheme has
incorporated elements of the NF approach~(as the leading
contribution) and the hard-scattering approach~(as the sub-leading
corrections), and provides a powerful and systematical means to
compute the radiative~(sub-leading nonfactorizable) corrections to
the NF approximation for the hadronic matrix elements. In
particular, the strong phases, which are very important for
studying the $CP$ violation in $B$-meson decays, are calculable
from the first principle. Detailed proofs and arguments can be
found in Ref.~\cite{bbns1}, and current status and recent
developments of this approach have also been reviewed recently
in~\cite{Du:2005vc}.

In a recent work~\cite{Li:2005wx}, we have studied the higher
order penguin contractions of spectator-scattering amplitudes
induced by the $b\to D g^\ast g^\ast$ transitions~(where $D=d$ or
$s$, depends on the specific decay mode, and the off-shell gluons
$g^\ast$ are either emitted from the internal quark loops,
external quark lines, or splitted off the virtual gluon of the
penguin diagrams), and investigated their impacts on the
$CP$-averaged branching ratios and $CP$-violating asymmetries of
$B\to \pi\pi, \pi K$ decays. It has been found that these higher
order penguin contraction contributions are not negligible in
two-body hadronic $B$-meson decays, particularly in the
penguin-dominated $B\to \pi K$ decays. Thus, combining the
findings in the literature \cite{hou1,simma,greub,bosch}, it would
be worthy to take into account these higher order penguin
contraction contributions to the exclusive hadronic $B$-meson
decays. This encourages us to further investigate their impacts on
the hadronic charmless $B \to PV$~(where $P$ and $V$ denote
pseudoscalar and vector mesons, respectively) decays.

$B\to PV$  decays are closely related to their $PP$ counterparts
because of their similar flavor structures, however, these modes
have apparent advantages in some cases. For example, due to the
less penguin pollution, $B\to \pi \rho$ decay modes are more
suitable than $B\to \pi \pi$ ones for extracting the weak angle
$\alpha$ of the unitarity triangle of the
Cabibbo-Kobayshi-Maskawa~(CKM) matrix~\cite{ckm}. Studies on
two-body hadronic $B\to PV$ decays are therefore very helpful to
deepen our understandings of the rare hadronic $B$-meson decays.
Earlier theoretical studies on $B{\to}PV$ decays based on various
approaches can be found, for example, in
Refs.~\cite{Ali:1998eb,Deshpande:1997rr}. With the accumulation of
new experimental data and the theoretical improvements, these
$B{\to}PV$ decay modes have also been reanalyzed
recently~\cite{Chiang:2003pm,Yang:2000xn,Du:2002up,bbns3}. In this
paper, we will reexamine these hadronic charmless $B\to PV$ decays
within the framework of QCDF approach and take into account the
higher order penguin contractions of spectator-scattering
amplitudes as mentioned above. Here we do not consider the decay
modes with an $\eta$ or $\eta^{\prime}$ meson in the final states,
since in this case there are  many additional unknown parameters
pertaining to these two particles, such as their contents,  mixing
angles, and the anomaly $g-g-\eta^{(\prime)}$ coupling, which
would  hinder us from getting reliable theoretical predictions.

This paper is organized as follows. Sec.~\ref{sec2} is devoted to
the theoretical framework. In this section, we first give the
relevant formulas describing the decay amplitudes of hadronic
$B\to PV$ decays at next-to-leading order in $\alpha_s$, and then
take into account contributions of the higher order penguin
contractions of spectator-scattering amplitudes induced by the
$b\to D g^\ast g^\ast$ transitions. In Sec.~\ref{sec3}, we give
our numerical results for $CP$-averaged branching ratios and
$CP$-violating asymmetries,  and discuss the impacts of the higher
order corrections on these quantities.  Detailed analysis of the
interesting decays $B\to \pi K^{\ast}$ and $B\to K \rho$, are also
presented in this section. Finally, we conclude with a summary in
Sec.~\ref{sec4}. Some useful functions and the input parameters
used in this paper are collected in Appendix A and B,
respectively.

\section{Theoretical framework for $B\to PV$ decays}
\label{sec2}

\subsection{The effective Hamiltonian for hadronic $B$-meson
decays}\label{sec2.1}

Using the operator product expansion~(OPE) and the renormalization
group equation~(RGE), the low energy effective Hamiltonian for
hadronic charmless $B$-meson decays in the SM can be written
as~\cite{Buchalla:1996vs}
\begin{equation}\label{Heff}
   {\cal H}_{\rm eff} = \frac{G_F}{\sqrt2} \sum_{p=u,c} \!
   \lambda_p^{(\prime)} \bigg( C_1\,Q_1^p + C_2\,Q_2^p
   + \!\sum_{i=3,\dots, 10}\! C_i\,Q_i + C_{7\gamma}\,Q_{7\gamma}
   + C_{8g}\,Q_{8g} \bigg) + \mbox{h.c.} \,,
\end{equation}
where $\lambda_p=V_{pb}\,V_{ps}^*$ (for $b\to s$ transition) and
$\lambda_p^\prime=V_{pb}\,V_{pd}^*$ (for $b\to d$ transition) are
products of the CKM matrix elements, and the unitarity relation
$-\lambda_t^{(\prime)}=\lambda_u^{(\prime)}+\lambda_c^{(\prime)}$
has been used. The effective operators, $Q_{i}$, governing a given
decay process, can be expressed explicitly as follows.
\begin{itemize}
\item[]{(i)} Current-current operators:
\begin{eqnarray}
   Q_1^p = (\bar p b)_{V-A} (\bar D p)_{V-A} \,,
    \hspace{2.5cm}
   Q^p_2 = (\bar p_i b_j)_{V-A} (\bar D_j p_i)_{V-A} \,,
\end{eqnarray}
\item[]{(ii)} QCD-penguin operators:
\begin{eqnarray}
   Q_3 = (\bar D b)_{V-A} \sum_{q}\,(\bar q q)_{V-A} \,,
    \hspace{1.7cm}
   Q_4 = (\bar D_i b_j)_{V-A} \sum_{q}\,(\bar q_j q_i)_{V-A} \,,
    \nonumber\\
   Q_5 = (\bar D b)_{V-A} \sum_{q}\,(\bar q q)_{V+A} \,,
    \hspace{1.7cm}
   Q_6 = (\bar D_i b_j)_{V-A} \sum_{q}\,(\bar q_j q_i)_{V+A} \,,
\end{eqnarray}
\item[]{(iii)} Electroweak penguin operators:
\begin{eqnarray}
   Q_7 = (\bar D b)_{V-A} \sum_{q}\,{\textstyle\frac32} e_q
    (\bar q q)_{V+A} \,, \hspace{1.11cm}
   Q_8 = (\bar D_i b_j)_{V-A} \sum_{q}\,{\textstyle\frac32} e_q
    (\bar q_j q_i)_{V+A} \,, \nonumber \\
   Q_9 = (\bar D b)_{V-A} \sum_{q}\,{\textstyle\frac32} e_q
    (\bar q q)_{V-A} \,, \hspace{0.98cm}
   Q_{10} = (\bar D_i b_j)_{V-A} \sum_{q}\,{\textstyle\frac32} e_q
    (\bar q_j q_i)_{V-A} \,,
\end{eqnarray}
\item[]{(iv)} Electro- and chromo-magnetic dipole operators:
\begin{eqnarray}
   Q_{7\gamma} = \frac{-e}{8\,\pi^2}\,m_b\,
    \bar D \,\sigma_{\mu\nu}\,(1+\gamma_5)\, F^{\mu\nu} b \,,
    \hspace{0.81cm}
   Q_{8g} = \frac{-g_s}{8\,\pi^2}\,m_b\,
    \bar D \,\sigma_{\mu\nu}\,(1+\gamma_5)\, G^{\mu\nu} b \,,
\end{eqnarray}
\end{itemize}
where $(\bar q_1 q_2)_{V\pm A}=\bar
q_1\gamma_\mu(1\pm\gamma_5)q_2$, $i,j$ are colour indices, $e_q$
is the quark electric charge in units of $|e|$, and a summation
over $q=u, d, s, c, b$ is implied. For $b\to d$ transition induced
decay modes, $D=d$, while for $b\to s$ transition induced ones,
$D=s$.

The Wilson coefficients $C_{i}(\mu)$ in Eq.~(\ref{Heff}) represent
all the physics contributions higher than the scale $\mu\sim {\cal
O}(m_{b})$. Numerical results for these coefficients evaluated at
different scales can be found in Ref.~\cite{Buchalla:1996vs}.

\subsection{Decay amplitudes at next-to-leading order in $\alpha_s$}
\label{sec2.2}

\begin{figure}[t]
\epsfxsize=9cm \centerline{\epsffile{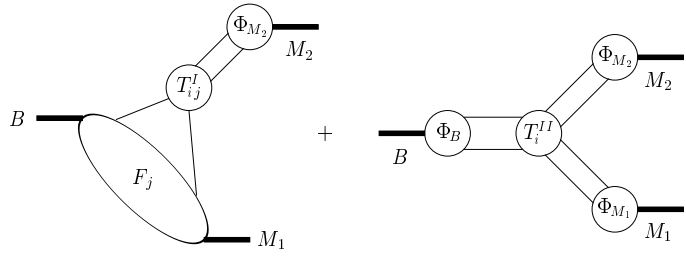}}
\centerline{\parbox{16cm}{\caption{\label{fact-fig} Graphical
representation of the factorization formula. Only one of the two
form-factor terms in Eq.~(\ref{fact}) is shown for simplicity.}}}
\end{figure}

With the low energy effective Hamiltonian given by
Eq.~(\ref{Heff}), the decay amplitude for a general hadronic
charmless $B\to PV$ decay can be written as
\begin{equation}\label{fac1}
   \langle PV|{\cal H}_{\rm eff}|B\rangle
   = \frac{G_F}{\sqrt2} \sum_{p=u,c} \lambda_p^{(\prime)}\,C_i\,
   \langle PV|Q_i^p|B\rangle \,.
\end{equation}
Then, the most essential theoretical problem in the calculation of
the decay amplitude resides in the evaluation of the hadronic
matrix elements of the local operators contained in the effective
Hamiltonian, $\langle PV|Q_i^p|B\rangle$. With the QCDF approach,
they could be simplified to a large extent. To leading power in
$\Lambda_{\rm QCD}/m_b$, but to all orders in perturbation theory,
these hadronic matrix elements obey the following factorization
formula~\cite{bbns1}
\begin{eqnarray}\label{fact}
\langle PV|Q_i^p|B\rangle &=&
 F_+^{B\to P}\,T_{V,i}^{\rm I}*f_V\,\Phi_V +
 A_0^{B\to V}\,T_{P,i}^{\rm I}*f_P\,\Phi_P \nonumber\\
 &&\mbox{} + T_i^{\rm II}*f_B\,\Phi_B*f_P\,\Phi_P*f_V\,\Phi_V \,,
\end{eqnarray}
where ${\Phi}_{M}$ are the LCDAs of the meson $M$, the $*$
products indicate convolutions of  the LCDAs and the
hard-scattering kernels $T^{\rm I,II}_{i}$. $F_+^{B\to P}$ and
$A_0^{B\to V}$ denote the heavy-to-light $B\to P$ and $B\to V$
transition form factors, respectively.  A graphical representation
of this formula is shown in Fig.~\ref{fact-fig}.

\begin{figure}[t]
\epsfxsize=10cm \centerline{\epsffile{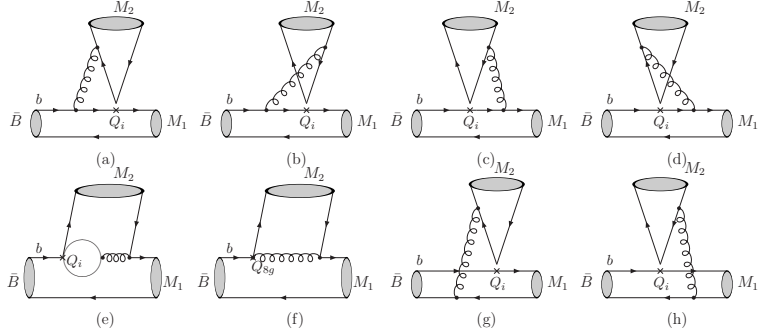}}
\centerline{\parbox{16cm}{\caption{\label{asfig} Order $\alpha_s$
corrections to the hard-scattering kernels $T_{M,i}^{\rm I}$
(coming from the diagrams (a)-(f)) and $T_i^{\rm II}$ (coming from
the last two diagrams).}}}
\end{figure}

When the power suppressed ${\cal O}(\Lambda_{\rm QCD}/m_b)$ terms
are neglected, $T^{\rm I,II}_{i}$  are dominated by hard gluon
exchanges, and hence calculable order by order in perturbative
QCD. The relevant Feynman diagrams contributing to these
hard-scattering kernels at next-to-leading order in $\alpha_s$ are
shown in Fig.~\ref{asfig}. The kernel $T_{M,i}^{\rm I}$ starts at
tree level and, at next-to-leading order in $\alpha_s$, contains
the sub-leading ``nonfactorizable" corrections coming from the
vertex-correction diagrams Figs.~\ref{asfig}(a-d) and the penguin
diagrams Figs.~\ref{asfig}(e-f). The kernel $T_i^{\rm II}$
contains the hard ``nonfactorizable" interactions between the
spectator quark and the emitted meson $M_{2}$. Its lowest order
contributions are of order $\alpha_s$ and can be depicted by the
hard spectator-scattering diagrams Figs.~\ref{asfig}(g-h). At
leading order, $T_{M,i}^{\rm I}=1$, $T_i^{\rm II}=0$, the QCDF
formula reproduce the NF results.

As stressed in Ref.~\cite{bbns2}, it should be borne in mind that
the factorization formula, Eq.~(\ref{fact}), does not imply that
the hadronic $B$-meson decays are perturbative in nature. Dominant
soft contributions to the decay amplitude do exist. However, all
these nonperturbative effects either are power-suppressed by
$\Lambda_{\rm QCD}/m_b$ or can be factorized into the transition
form factors and the meson LCDAs.

With the above discussions about the effective Hamiltonian for
hadronic $B$-meson decays and the QCDF formula for the hadronic
matrix element, the decay amplitude for a general hadronic
charmless $B\to PV$ decay, in the heavy quark limit, can then be
rewritten as
\begin{equation}\label{am1}
 {\cal A}(B\to PV) = \frac{G_{F}}{\sqrt{2}}\,\sum_{p=u,c}\,
 \sum_{i=1}^{10}\, \lambda_{p}^{(\prime)}\, a_{i}^{p}\,
 {\langle}PV{\vert}Q_{i}{\vert}B{\rangle}_{F}\,,
\end{equation}
where ${\langle}PV{\vert}Q_{i}{\vert}B{\rangle}_{F}$ is the
factorized hadronic matrix element, which has the same definition
as that in the NF approach. All the ``nonfactorizable'' effects
are encoded in the coefficients $a_{i}^p$, which are process
dependent and can be calculated perturbatively. Following Beneke
\textsl{et al.}~\cite{bbns3}, the general form of the coefficients
$a_i^p$~($i=1,...,10$) at next-to-leading order in $\alpha_s$,
with $M_1$ being the meson picking up the spectator quark and
$M_2$ the emitted meson, can be written as
\begin{equation} \label{ai}
a_i^p (M_1 M_2) =(C_i + \frac{{C_{i \pm 1} }}{{N_c }})N_i (M_2 )+
\frac{{C_{i \pm 1} }}{{N_c }}\frac{{C_F \alpha _s
}}{{4\pi}}\left[V_i (M_2 ) + \frac{{4\pi ^2 }}{{N_c }}H_i (M_1 M_2
)\right] + P_i^p (M_2)\,,
\end{equation}
where the upper (lower) signs apply when $\it{i}$ is odd (even).
The quantities $V_i(M_2)$ account for one-loop vertex corrections,
$H_i (M_1 M_2)$ for hard-spectator interactions, and $P_i^p (M_2
)$ for penguin contributions. Explicit expressions for these
quantities can be found in Ref.~\cite{bbns3}.

It is noted that, in calculations of the decay amplitudes for
hadronic charmless $B$-meson decays, the coefficients
$a_i^p$~($i=3,...,10$) always appear in pairs. So, for the
two-body hadronic charmless $B\to PV$ decays, one can define the
following quantities $\alpha_i^p$ in terms of the coefficients
$a_i^p$ defined in Eq.~(\ref{ai})~\cite{bbns3}
\begin{eqnarray}\label{ais}
   \alpha_1(M_1 M_2) &=& a_1(M_1 M_2) \,, \nonumber\\
   \alpha_2(M_1 M_2) &=& a_2(M_1 M_2) \,, \nonumber\\
   \alpha_3^p(M_1 M_2) &=& \left\{
    \begin{array}{cl}
     a_3^p(M_1 M_2) - a_5^p(M_1 M_2) \,;
      & \quad \mbox{if~} M_1 M_2=VP \,, \\
     a_3^p(M_1 M_2) + a_5^p(M_1 M_2) \,;
      & \quad \mbox{if~} M_1 M_2=PV  \,,
    \end{array}\right. \nonumber\\
   \alpha_4^p(M_1 M_2) &=& \left\{
    \begin{array}{cl}
     a_4^p(M_1 M_2) + r_{\chi}^{M_2}\,a_6^p(M_1 M_2) \,;
      & \quad \mbox{if~} M_1 M_2=PV \,, \\
     a_4^p(M_1 M_2) - r_{\chi}^{M_2}\,a_6^p(M_1 M_2) \,;
      & \quad \mbox{if~} M_1 M_2=VP\,,
    \end{array}\right.\\
   \alpha_{3,ew}^p(M_1 M_2) &=& \left\{
    \begin{array}{cl}
     a_9^p(M_1 M_2) - a_7^p(M_1 M_2) \,;
      & \quad \mbox{if~} M_1 M_2=VP \,, \\
     a_9^p(M_1 M_2) + a_7^p(M_1 M_2) \,;
      & \quad \mbox{if~} M_1 M_2=PV  \,,
    \end{array}\right. \nonumber\\
   \alpha_{4,ew}^p(M_1 M_2) &=& \left\{
    \begin{array}{cl}
     a_{10}^p(M_1 M_2) + r_{\chi}^{M_2}\,a_8^p(M_1 M_2) \,;
      & \quad \mbox{if~} M_1 M_2=PV \,, \\
     a_{10}^p(M_1 M_2) - r_{\chi}^{M_2}\,a_8^p(M_1 M_2) \,;
      & \quad \mbox{if~} M_1 M_2=VP\,,
     \end{array}\right.\nonumber
\end{eqnarray}
with the scale-dependent ratio $r_\chi^{M_2}$ defined as
\begin{equation}\label{chiral factor}
   r_\chi^P(\mu) = \frac{2\,m_P^2}{m_b(\mu)(m_{q_1}+m_{q_2})(\mu)},\qquad\,
   r_\chi^V(\mu) = \frac{2\,m_V}{m_b(\mu)}\,\frac{f_V^\perp(\mu)}{f_V}\,,
\end{equation}
where all quark masses are running current masses defined in the
$\overline{\rm MS}$ scheme, and $f_V^\perp(\mu)$ is the
scale-dependent transverse decay constant of vector meson.
Although all these terms  proportional to $r_\chi^{M_2}$ are
formally power suppressed by $\Lambda_{\rm QCD}/m_b$ in the
heavy-quark limit, they are not small numerically. In particular,
the factor $r_\chi^P(\mu)$ is chirally enhanced and important for
charmless $B$ decays~\cite{bbns2,bbns3}.

\begin{figure}[t]
\epsfxsize=11cm \centerline{\epsffile{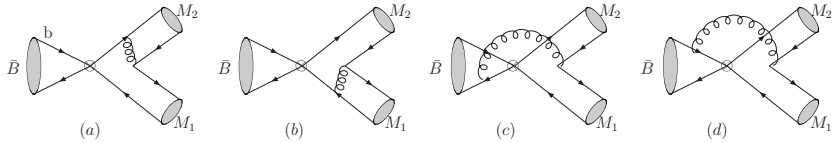}}
\centerline{\parbox{16cm}{\caption{\label{annhfig} The weak
annihilation diagrams of order $\alpha_s$.}}}
\end{figure}

According to the arguments in \cite{bbns1}, the weak annihilation
contributions to the decay amplitudes are power suppressed, and
hence do not appear in the QCDF formula, Eq.~(\ref{fact}).
Nevertheless, these contributions may be numerically important for
realistic $B$-meson decays. At order $\mathcal{O}(\alpha_s)$, the
annihilation kernels arise from the four diagrams shown in
Fig.~\ref{annhfig}. They result in a further contribution to the
hard scattering kernel $T_i^{\rm II}$ in the QCDF formula,
Eq.(\ref{fact}). However, within the QCDF formalism, these
annihilation topologies violate factorization because of the
end-point divergence. In this work, following the treatment of
Refs.~\cite{bbns2,Feldmann:2004mg}, we will introduce a cutoff to
parameterize these contributions and express the weak annihilation
decay amplitudes as
\begin{equation} \label{am2}
{\cal A}^{ann}(B\to
PV)\propto\frac{G_F}{\sqrt{2}}\sum_{p=u,c}\sum_{i}\lambda_p^{(\prime)}
f_B\,f_{M_1}\,f_{M_2}\,b_i(M_{1}M_{2})\,,
\end{equation}
where $f_B$ and $f_M$ are the decay constants of the initial $B$
and the final-state mesons, respectively. The parameters
$b_i(M_{1}M_{2})$ describe the annihilation contributions and
their explicit expressions can be found in
Refs.~\cite{bbns2,bbns3}.

The explicit expressions of the decay amplitudes for hadronic
charmless $B\to PV$ decays, including the weak annihilation
contributions, can be found, for example, in
Refs.~\cite{Du:2002up,bbns3}. It should be noted that, within the
QCDF framework, all the ``nonfactorizable" power suppressed
contributions except for the hard spectator interactions and weak
annihilation contributions are neglected. In addition, in the
evaluation of the hard spectator and weak annihilation terms, the
running coupling constant and the Wilson coefficients should be
evaluated at an intermediate scale $\mu_h\sim(\Lambda_{\rm
QCD}\,m_b)^{1/2}$ rather than the scale $\mu\sim m_b$.  However,
the evolution of $C_{i}(\mu)$  down to $\mu_{h}$ is highly
nontrivial, since the RGE will change below the scale $m_{b}$. To
deal with this problem, one may have to turn  to SCET which is the
appropriate effective theory for QCD below the $m_{b}$ scale.
However, in this paper, we restrict ourself to QCDF and adopt the
treatments of evolution of the $C_{i}(\mu)$ as done in
\cite{bbns2}, i.e., we do not take into account the charm and
bottom threshold and evolve the Wilson coefficients in a
5-flavored theory. With this approximation, in the evolution of
the Wilson coefficients, all logs of the form $\log{\mu/M_W}$ have
been summed, while logs of the form $\log{\mu/m_b}$ and
$\log{\mu/m_c}$ are not. Since the latter two terms are never
large with $\mu\geq m_b/2$,  the approximation would work  to  the
precision in this paper.~\footnote{We thank M.~Beneke for pointing
out this point to us.} Specifically, we shall use
$\mu_h=\sqrt{\Lambda_h\,\mu}$ with $\Lambda_h=0.5~{\rm GeV}$ in
our numerical calculations.

\subsection{Penguin contractions of spectator-scattering amplitudes and their contributions to $B\to PV$  decays}\label{sec2.3}

\begin{figure}[t]
\epsfxsize=9cm \centerline{\epsffile{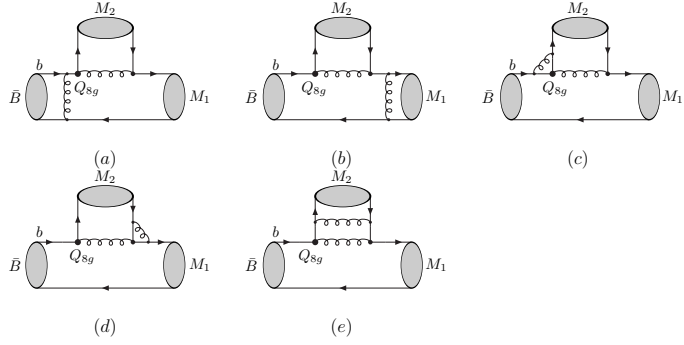}}
\centerline{\parbox{16cm}{\caption{\label{fig:factorizable} \small
Representative Feynman diagrams induced by the $b\to D g^\ast
g^\ast$ transitions which are not needed to evaluate.  Only the
chromo-magnetic dipole operator $Q_{8g}$ contributions are shown.
With the operator $Q_{8g}$ replaced by the other operators, the
corresponding Feynman diagrams can also be obtained.}}}
\end{figure}

\begin{figure}[t]
\epsfxsize=9cm \centerline{\epsffile{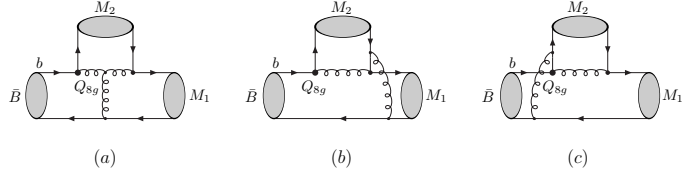}}
\centerline{\parbox{16cm}{\caption{\label{Q8gfig} \small
Chromo-magnetic dipole operator $Q_{8g}$ contributions induced by
the $b\to D g^\ast g^\ast$ transitions.}}}
\end{figure}

\begin{figure}[t]
\epsfxsize=9cm \centerline{\epsffile{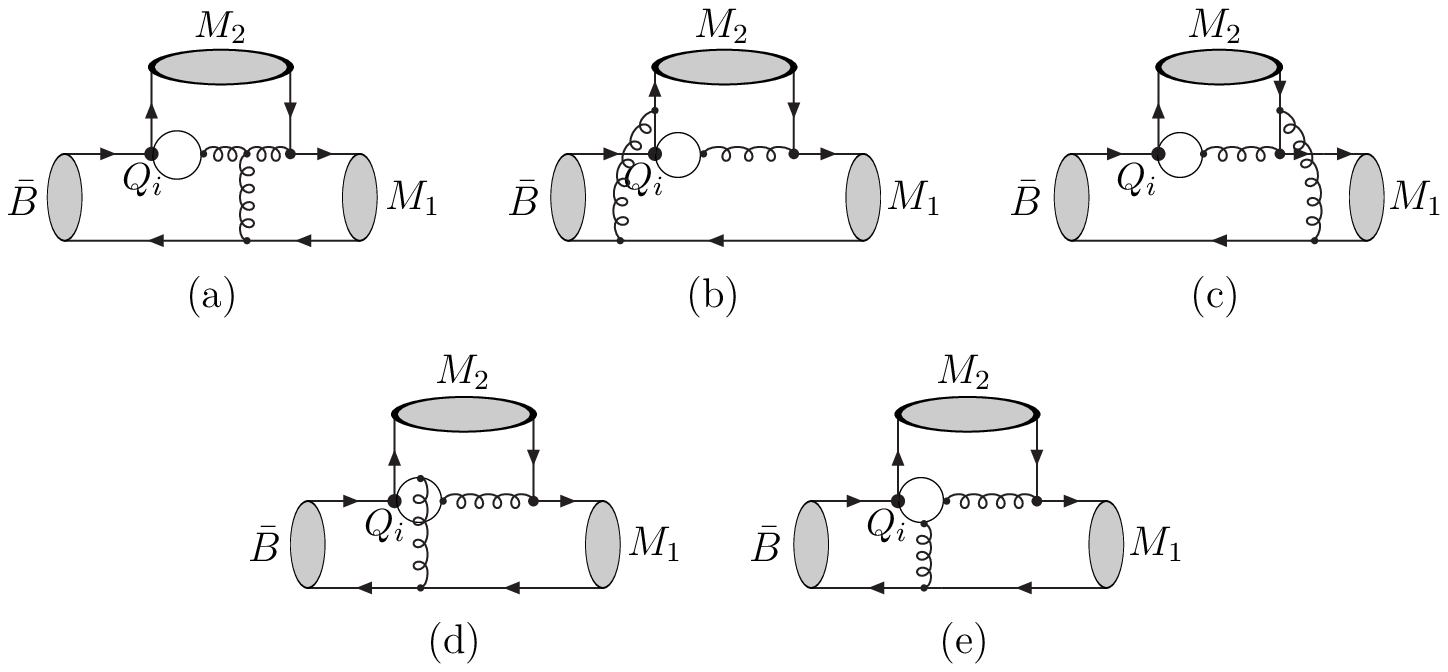}}
\centerline{\parbox{16cm}{\caption{\label{penguinfig} \small
Penguin operator $Q_i$ contributions induced by the $b\to D g^\ast
g^\ast$ transitions.}}}
\end{figure}

At the quark level, the $b \to D g^\ast g^\ast$ transitions can
occur in many different manners as depicted by
Figs.~\ref{fig:factorizable}--\ref{penguinfig}. For example, one
of the two off-shell gluons can radiate from the external quark
line, while the other one comes from the chromo-magnetic dipole
operator $Q_{8g}$ as Figs.~\ref{Q8gfig}(b) and~\ref{Q8gfig}(c) or
from the internal quark loops of the penguin diagrams as
Figs.~\ref{penguinfig}(b) and~\ref{penguinfig}(c). On the other
hand, the two off-shell gluons can also radiate from the internal
quark loops as Figs.~\ref{penguinfig}(d) and~\ref{penguinfig}(e)
or split off the virtual gluon of the penguin diagrams as
Figs.~\ref{Q8gfig}(a) and~\ref{penguinfig}(a). Here we do not
consider the Feynman diagrams of the category shown in
Fig.~\ref{fig:factorizable}, since their contributions can be
absorbed into the definitions of heavy-to-light transition form
factors as Figs.~\ref{fig:factorizable}(a)
and~\ref{fig:factorizable}(b) and the meson LCDAs as
Figs.~\ref{fig:factorizable}(e), or are further suppressed by
$\frac{1}{16\pi^2}$ as Figs.~\ref{fig:factorizable}(c)
and~\ref{fig:factorizable}(d). It is easy to clarify this point by
comparing the strength of Fig.~\ref{fig:factorizable}(c) to that
of Fig.~\ref{Q8gfig}(a).


As shown by Figs.~\ref{Q8gfig} and~\ref{penguinfig}, these Feynman
diagrams should be the dominant sources contributing to the
penguin contractions of spectator-scattering amplitudes of order
$\alpha_s^2$, \textit{since they are not two-loop QCD diagrams,
and hence there are no additional $\frac{1}{16 \pi^2}$ suppression
factor in their contributions compared to the genuine two-loop
ones of order $\alpha_s^2$}. Studying these contributions could be
very helpful for our understandings of the higher order
perturbative corrections to the rare hadronic $B$-meson decays
within the QCDF formalism.

We start with the calculations of the Feynman diagrams in
Fig.~\ref{Q8gfig}. In this case, the $b$ quark weak decay is
induced by the chromo-magnetic dipole operator $Q_{8g}$, and the
calculation is straightforward with the result given by
\begin{eqnarray}\label{AQ8g}
{\cal A}_{Q_{8g}}&=& i\,\frac{\alpha_s^2\, f_B \,f_{M_1}\,
f_{M_2}}{N_c^3}\,\lambda_t^{(\prime)}\,\int_0^1\!
d{\xi}\,\frac{\Phi_1^B (\xi)}{\xi}\,\nonumber\\
&& \times\,\int_0^1\!dudv\,\left\{
\Phi_{M_2}(u)\,\Phi_{M_1}(v)\,\biggl[\frac{3\,( 3 - v )}{2\,( 1 -
u )\,( 1 - v ) \,v} + \frac{1}{6\,( 1 - u )\,( 1 -
v)}\,\biggl]\,\right. \nonumber\\
&& \qquad + \left.
r_\chi^{M_1}\,\Phi_{M_2}(u)\,\Phi_{m_1}(v)\,\biggl[\frac{ 3\,( 3 -
u - v + u\,v )}{ 2\,( 1 - u )^2\,( 1 - v )\,v} + \frac{ 2 -
u}{6\,( 1 - u ) \, u\,( 1 - v )} \,\biggl] \right. \nonumber\\
&& \qquad + \left.
r_\chi^{M_1}\,r_\chi^{M_2}\,\Phi_{m_2}(u)\,\Phi_{m_1}(v)\,\biggl[\frac{1}{6\,
(1 - u ) \,( 1 - v ) } + \frac{3\,( 3 - v ) } {2\,( 1 - u )\,( 1 -
v )\,v}\,\biggl] \right. \nonumber\\
&& \qquad + \left.
r_\chi^{M_2}\,\Phi_{m_2}(u)\,\Phi_{M_1}(v)\,\biggl[\frac{3\,( 3 -
u - v - u\,v ) } {2\,( 1 - u ) \,( 1 - v ) \,v} + \frac{ 1 +
u}{6\,( 1 - u ) \,( 1 - v ) }\,\biggl] \,\right\},
\end{eqnarray}
when $M_1$ is a pseudoscalar and $M_2$ a vector meson. For the
opposite case of a vector $M_1$ and a pseudoscalar $M_2$, one
needs only change the signs of the last two terms in the bracket
of Eq.~(\ref{AQ8g}). Here $\lambda_t=V_{tb}V_{ts}^\ast$  and $\lambda_t^\prime=V_{tb}V_{td}^\ast$
 are products of the CKM matrix elements, $\Phi_M$ and $\Phi_m$ denote the
leading-twist and twist-3 LCDAs of the meson $M$ in the final
states, respectively. In our calculation, we use the LCDAs in the asymptotic limit
\begin{equation}\label{DA}
\Phi_{P}(x)=\Phi_{V}(x)=6x(1-x),\qquad \Phi_{p}(x)=1, \qquad
\Phi_{v}=3(2x-1),
\end{equation}
and have neglected the tree-particle LCDAs and deviations from the
asymptotic limit.

In calculating the Feynman diagrams in Fig.~\ref{penguinfig}, we
adopt the method proposed by Greub and Liniger~\cite{greub}. We
first calculate the fermion loops in these individual Feynman
diagrams, and then insert these building blocks into the entire
Feynman diagrams to obtain the final results. In evaluating the
internal quark loop diagrams, we shall adopt the naive dimensional
regularization~(NDR) scheme and the modified minimal
subtraction~($\overline{\rm MS}$) scheme. In addition, we shall
adopt Feynman gauge for the gluon propagator
throughout this paper. The gauge invariance will be  guaranteed when the full set of Feynman diagrams 
are summed with the external quarks on-mass-shell~\cite{Cheng:1999gs}.  However,  
we must be care of the gauge dependence in our calculation,  since only a subset   $\mathcal{O}(\alpha^{2}_{s})$ Feynman diagrams are calculated.  After careful checking, we find that each  Feynman diagram in Fig.5. and 6 is  gauge independent.  The detail checking can be found in Appendix C.    Analogous to the calculation of
the penguin diagrams in Fig.~\ref{asfig}(e), we should also take
into account the two distinct penguin contractions of the
four-quark operators in the weak interaction vertex .

\begin{figure}[t]
\epsfxsize=9cm \centerline{\epsffile{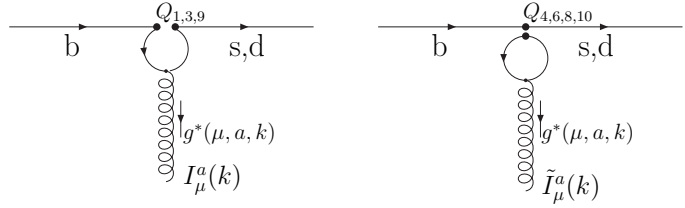}}
\centerline{\parbox{16cm}{\caption{\label{buildingblock1} \small
Building blocks $I_\mu^a(k)$ (corresponding to contractions of the
operators $Q_{1,3,9}$) and $\tilde{I}_\mu^a(k)$ (corresponding to
contractions of the operators $Q_{4,6,8,10}$) for
Figs.~\ref{penguinfig}(a)--\ref{penguinfig}(c).}}}
\end{figure}

As shown in Fig.~\ref{penguinfig}, the first three Feynman
diagrams have the same building block $I_\mu^a(k)$ (corresponding
to contractions of the operators $Q_{1,3,9}$) or
$\tilde{I}_\mu^a(k)$ (corresponding to contractions of the
operators $Q_{4,6,8,10}$). These building blocks can be depicted
by Fig.~\ref{buildingblock1} and given by
\begin{eqnarray}
I_\mu^a(k) &=& \frac{g_s}{4\,\pi^2}\,\Gamma(\frac{\epsilon}{2})\,
(2-\epsilon)\,(4\pi\mu^2)^{\frac{\epsilon}{2}}\,(k_\mu\,\kslash-k^2
\gamma_\mu)\,(1-\gamma_5)\,T^a\,\nonumber\\
&& \mbox\,
\times\int_0^1\!dx\,
\frac{x\,(1-x)}{\left[m_q^2-x(1-x)\,k^2-i\,\delta\,\right]^{\frac{\epsilon}{2}}}\,,\\
\tilde{I}_\mu^a(k) &=&
\frac{g_s}{2\,\pi^2}\,\Gamma(\frac{\epsilon}{2})\,
(4\pi\mu^2)^{\frac{\epsilon}{2}}\,(k_\mu\,\kslash-k^2\gamma_\mu)\,
(1-\gamma_5)\,T^a\,\nonumber\\
&& \mbox\, \times\int_0^1\!dx\,
\frac{x\,(1-x)}{\left[m_q^2-x(1-x)\,k^2-i\,\delta\,\right]^{\frac{\epsilon}{2}}}\,,
\end{eqnarray}
where $k$ is the momentum of the off-shell gluon,
$T^a=\frac{\lambda^a}{2}$, with $\lambda^a$ the Gell-Mann
matrices, $g_s$ is the strong coupling constant, and $m_q$ the
pole mass of the quark propagating in the fermion loops. We have
used $d=4-\epsilon$. After performing the subtraction with the
$\overline{\rm MS}$ scheme, we get
\begin{eqnarray}\label{Ibuilding}
I_\mu^a(k) &=&-\frac{g_s}{8\,\pi^2}\left[\frac{2}{3}-\frac{4}{3}\,
\ln \frac{\mu}{m_b} - G(s_q,r)\,\right] (k_\mu\,\kslash-k^2
\gamma_\mu)\,(1-\gamma_5)\,T^a\,,\\
\tilde{I}_\mu^a(k) &=& -\frac{g_s}{8\,\pi^2}\left[-\frac{4}{3}\,
\ln \frac{\mu}{m_b} - G(s_q,r)\,\right] (k_\mu\,\kslash-k^2
\gamma_\mu)\,(1-\gamma_5)\,T^a\,,
\end{eqnarray}
with the function $G(s_q,r)$ defined by
\begin{equation}\label{Gfunction}
G(s_q,r) = -4\int_0^1\!dx\,x\,(1-x) \ln[s_q-x(1-x)r-i\delta\,],
\end{equation}
where $s_q=m_q^2/m_b^2$, $r=k^2/m_b^2$, and the term $i\delta$ is
the ``$\epsilon$-prescription". The free indices $\mu$ and $a$
should be contracted with the gluon propagator when inserting
these building blocks into the entire Feynman diagrams.

\begin{figure}[t]
\epsfxsize=8cm \centerline{\epsffile{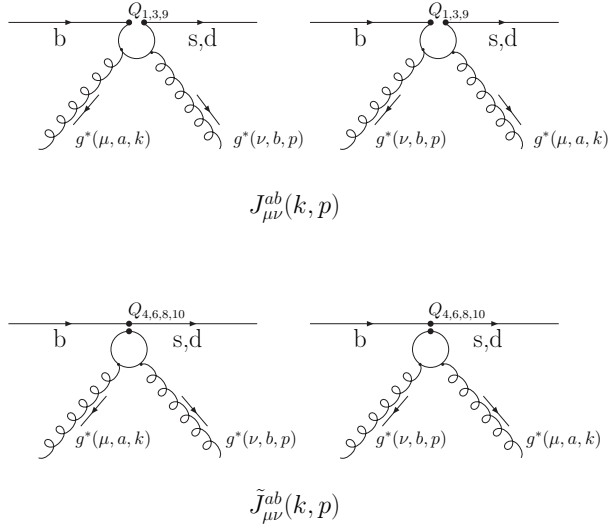}}
\centerline{\parbox{16cm}{\caption{\label{buildingblock2} \small
Building blocks $J_{\mu\nu}^{ab}(k,p)$ (corresponding to
contractions of the operators $Q_{1,3,9}$) and
$\tilde{J}_{\mu\nu}^{ab}(k,p)$ (corresponding to contractions of
the operators $Q_{4,6,8,10}$) for Figs.~\ref{penguinfig}(d)
and~\ref{penguinfig}(e).}}}
\end{figure}

The sum of the fermion loops in the last two diagrams in
Fig.~\ref{penguinfig} are denoted by the building block
$J_{\mu\nu}^{ab}(k,p)$ (corresponding to contractions of the
operators $Q_{1,3,9}$) or $\tilde{J}_{\mu\nu}^{ab}(k,p)$
(corresponding to contractions of the operators $Q_{4,6,8,10}$),
which are depicted by Fig.~\ref{buildingblock2}. Using the
decomposition advocated by \cite{simma,greub}, these building
blocks can be expressed as
\begin{eqnarray}
J^{ab}_{\mu\nu}(k,p) &=& T^{+}_{\mu\,\nu}(k,p)\,\Big\{ T^{a},
T^{b}\Big\} +T^{-}_{\mu\,\nu}(k,p)\Big[ T^{a}, T^{b} \Big]\,,\\
\tilde{J}^{ab}_{\mu\nu}(k,p) &=&
\tilde{T}^{+}_{\mu\,\nu}(k,p)\,\Big\{ T^{a}, T^{b}\Big\}
+\tilde{T}^{-}_{\mu\,\nu}(k,p)\Big[ T^{a}, T^{b} \Big]\,,
\end{eqnarray}
where the first~(second) part is symmetric~(antisymmetric) with
respect to the color structures of the two off-shell gluons. Here
$k~(p)$, $a~(b)$, and $\mu~(\nu)$ are the momentum, color, and
polarization of the off-shell gluons, respectively. Below we refer
to the gluon with indices $(\nu,b,p)$ as the one connected to
the spectator quark.

In the NDR scheme, after performing the (shifted) loop momentum
integration, we can represent the quantities
$T^{\pm}_{\mu\,\nu}(k,p)$ and $\tilde{T}^{\pm}_{\mu\,\nu}(k,p)$
as~\cite{simma,greub}
\begin{eqnarray}
T^{+}_{\mu\,\nu}(k,p)&=& \frac{\alpha_s}{4\,\pi} \,\biggl[
E(\mu,\nu,k)\,\Delta i_{5}\, +E(\mu,\nu, p)\,\Delta i_{6}\,
-E(\mu, k, p)\, \frac{k_{\nu}}{k\cdot p}\,\Delta i_{23}\,\biggl. \nonumber\\
&&\mbox\,\biggl. - E(\mu, k, p)\, \frac{p_{\nu}}{k\cdot p}\,\Delta
i_{24}\, -E(\nu, k, p)\, \frac{k_{\mu}}{k\cdot p}\, \Delta i_{25}
-E(\nu, k, p)\, \frac{p_{\mu}}{k\cdot p}\, \Delta i_{26} \,
\biggl]\,L\,\label{TPfunction},\\
T^{-}_{\mu\,\nu}(k,p)&=& \frac{\alpha_s}{4\,\pi}\, \biggl[ \kslash
\,g_{\mu\nu}\, \Delta i_{2}\, + \pslash \,g_{\mu\nu}\, \Delta
i_{3}\, + \gamma_{\mu}\, k_{\nu}\, \Delta i_{8}\, +\gamma_{\mu}\,
p_{\nu}\, \Delta i_{9}\, +\gamma_{\nu}\, k_{\mu}\, \Delta i_{11}\,
\biggl.\nonumber \\
&&\mbox \,\biggl. +\gamma_{\nu}\, p_{\mu}\, \Delta i_{12}\,
+\kslash\,\frac{ k_{\mu}k_{\nu} }{k\cdot p}\, \Delta i_{15}\,
+\kslash\,\frac{ k_{\mu}p_{\nu} }{k\cdot p}\, \Delta i_{16}\,
+\kslash\,\frac{ p_{\mu}k_{\nu} }{k\cdot p}\, \Delta i_{17}\,
+\kslash\,\frac{ p_{\mu}p_{\nu} }{k\cdot p}\, \Delta i_{18}\,
\biggl. \nonumber \\
&&\mbox \,\biggl. +\pslash\,\frac{ k_{\mu}k_{\nu} }{k\cdot p}\,
\Delta i_{19}\, +\pslash\,\frac{ k_{\mu}p_{\nu} }{k\cdot p}\,
\Delta i_{20}\, +\pslash\,\frac{ p_{\mu}k_{\nu} }{k\cdot p}\,
\Delta i_{21}\, +\pslash\,\frac{ p_{\mu}p_{\nu} }{k\cdot p}\,
\Delta i_{22}\,\biggl]\,L\,\label{TMfunction},\\
\tilde{T}^{+}_{\mu\,\nu}(k,p)&=& a\, T^{+}_{\mu\,\nu}(k,p)\,
\label{TPtfunction},\\
\tilde{T}^{-}_{\mu\,\nu}(k,p)&=& T^{-}_{\mu\,\nu}(k,p)\,
+\frac{\alpha_s}{4\,\pi}\, \biggl[ \kslash \,g_{\mu\nu}\,\frac43\,
-\pslash \,g_{\mu\nu}\, \frac43\, -\gamma_{\mu}\, k_{\nu}\,
\frac83\, -\gamma_{\mu}\, p_{\nu}\,\frac43\,  +\gamma_{\nu}\,
k_{\mu}\, \frac43\,\biggl. \nonumber \\
&&\mbox \,\biggl. + \gamma_{\nu}\, p_{\mu}\, \frac83\,
\biggl]\,L\,\label{TMtfunction},
\end{eqnarray}
where $L=1-\gamma_5$, and the matrix $E$ in Eq.~(\ref{TPfunction})
is defined by
\begin{eqnarray}
E(\mu,\nu,k) &=& \gamma_\mu \gamma_\nu \kslash\,-\gamma_\mu
k_\nu\,+\gamma_\nu k_\mu\,
-\kslash \, g_{\mu\,\nu}\,\nonumber\\
&=&
-i\,\epsilon_{\mu\nu\alpha\beta}\,k^\alpha\gamma^\beta\gamma_5\,,
\end{eqnarray}
with the second line obtained in a four dimensional context using
the Bjorken-Drell conventions. The parameter $a$ in
Eq.~(\ref{TPtfunction}) denotes the chiral structure of the
four-quark operators in the weak vertex with $a=\pm$ corresponding
to $(V-A) \otimes (V\mp A)$, respectively. Explicit expressions
for the dimensionally regularized $\Delta i$ functions can be
found in Appendix B of Ref.~\cite{Li:2005wx}.

Equipped with these building blocks, we can now evaluate all the
Feynman diagrams in Fig.~\ref{penguinfig}. After direct
calculations, the final results of these penguin contractions of
spectator-scattering amplitudes for hadronic charmless $B\to PV$
decays can be expressed as
\begin{eqnarray}
{\cal A}_{Q_{1}} &=& -i\,\frac{\alpha_s^2\, f_B\, f_{M_1}\,
f_{M_2}}{N_c^3}\,\lambda_p^{(\prime)}\,\int_0^1\!
d{\xi}\,\frac{\Phi_1^B(\xi)}{\xi}\,\int_0^1\!dudv\,\biggl\{
\left[\frac{2}{3}-\frac{4}{3}\, \ln{\frac{\mu}{m_b}}
- G(s_p,\bar u)\,\right]\,f_1(u,v)\,\biggl.\,\nonumber\\
&& \biggl.\, + \left[\frac{2}{3}-\frac{4}{3}\,\ln{\frac{\mu}{m_b}}
- G(s_p,\bar u\,v)\,\right]\,f_2(u,v)\, +
f_1(u,v,m_p)\,\biggl\}\,,\label{AQ1}\\
{\cal A}_{Q_{3}} &=& i\,\frac{\alpha_s^2\, f_B\, f_{M_1}\,
f_{M_2}}{N_c^3}\,\lambda_t^{(\prime)}\,\int_0^1\!
d{\xi}\,\frac{\Phi_1^B(\xi)}{\xi}\,\int_0^1\!dudv\,\biggl\{
\left[f_1(u,v,0) + f_1(u,v,1)\,\right]\,\biggl. \nonumber\\
&& \biggl.\, +
\left[\frac{4}{3}-\frac{8}{3}\,
\ln{\frac{\mu}{m_b}} - G(0,\bar u) - G(1,\bar u) \,\right]\,f_1(u,v)\,\biggl.\,\nonumber\\
&& \biggl.\, +
\left[\frac{4}{3}-\frac{8}{3}\,
\ln{\frac{\mu}{m_b}} - G(0,\bar u\,v) - G(1,\bar u\,v)\,\right]\,f_2(u,v)\,
\biggl\}\,,\label{AQ3}\\
{\cal A}_{Q_{9}} &=& -\frac{1}{2}\,{\cal A}_{Q_{3}}\,,\label{AQ9}\\
{\cal A}_{Q_{4}} &=& i\,\frac{\alpha_s^2\, f_B\, f_{M_1}\,
f_{M_2}}{N_c^3}\,\lambda_t^{(\prime)}\,\int_0^1\!
d{\xi}\,\frac{\Phi_1^B(\xi)}{\xi}\,\int_0^1\!dudv\,\biggl\{
\left[(n_f-2)\,f_2(u,v,0)\, + f_2(u,v,m_c) \right.\, \biggl. \nonumber\\
&& \biggl.\, \left.\, + f_2(u,v,m_b)\,\right]\, +
\left[-\frac{4\,n_f}{3}\,\ln \frac{\mu}{m_b}
- (n_f-2)\,G(0,\bar u)\, - G(s_c,\bar u)\, - G(1,\bar u)\,\right]\, f_1(u,v)\,\biggl. \nonumber\\
&& \biggl.\, + \left[-\frac{4\,n_f}{3}\,\ln \frac{\mu}{m_b}
- (n_f-2)\,G(0,\bar u\, v)\, - G(s_c,\bar u\, v)\, - G(1,\bar u\, v)\,\right]\, f_2(u,v)\,
\biggl\}\,,\label{AQ4}\\
{\cal A}_{Q_{6}} &=& i\,\frac{\alpha_s^2\, f_B\, f_{M_1}\,
f_{M_2}}{N_c^3}\,\lambda_t^{(\prime)}\,\int_0^1\!
d{\xi}\,\frac{\Phi_1^B(\xi)}{\xi}\,\int_0^1\!dudv\,\biggl\{
\left[(n_f-2)\,f_3(u,v,0)\, + f_3(u,v,m_c) \right.\, \biggl. \nonumber\\
&& \biggl.\, \left.\, + f_3(u,v,m_b)\,\right]\, +
\left[-\frac{4\,n_f}{3}\,\ln \frac{\mu}{m_b}
- (n_f-2)\,G(0,\bar u)\, - G(s_c,\bar u)\, - G(1,\bar u)\,\right]\, f_1(u,v)\,\biggl. \nonumber\\
&& \biggl.\, + \left[-\frac{4\,n_f}{3}\,\ln \frac{\mu}{m_b} -
(n_f-2)\,G(0,\bar u\, v)\, - G(s_c,\bar u\, v)\, - G(1,\bar u\,
v)\,\right]\, f_2(u,v)\,\biggl\}\,,\label{AQ6}\\
{\cal A}_{Q_{8}} &=& i\,\frac{\alpha_s^2\, f_B\, f_{M_1}\,
f_{M_2}}{N_c^3}\,\lambda_t^{(\prime)}\,\int_0^1\!
d{\xi}\,\frac{\Phi_1^B(\xi)}{\xi}\,\int_0^1\!dudv\,\biggl\{
\left[f_3(u,v,m_c) - \frac{1}{2}\,f_3(u,v,m_b)\,\right]\, \biggl. \nonumber\\
&& \biggl.\, +
\left[-\frac{2}{3}\,\ln \frac{\mu}{m_b}\, - G(s_c,\bar u)\, + \frac{1}{2}\,G(1,\bar u)\,\right]\, f_1(u,v)\,\biggl. \nonumber\\
&& \biggl.\, + \left[-\frac{2}{3}\,\ln \frac{\mu}{m_b}\, -
G(s_c,\bar u\,v)\, + \frac{1}{2}\,G(1,\bar u\,v)\,\right]\,
f_2(u,v)\, \biggl\}\,,\label{AQ8}\\
{\cal A}_{Q_{10}} &=& i\,\frac{\alpha_s^2\, f_B\, f_{M_1}\,
f_{M_2}}{N_c^3}\,\lambda_t^{(\prime)}\,\int_0^1\!
d{\xi}\,\frac{\Phi_1^B(\xi)}{\xi}\,\int_0^1\!dudv\,\biggl\{
\left[f_2(u,v,m_c) - \frac{1}{2}\,f_2(u,v,m_b)\,\right]\, \biggl. \nonumber\\
&& \biggl.\, +
\left[-\frac{2}{3}\,\ln \frac{\mu}{m_b}\, - G(s_c,\bar u)\, + \frac{1}{2}\,G(1,\bar u)\,\right]\, f_1(u,v)\,\biggl. \nonumber\\
&& \biggl.\, + \left[-\frac{2}{3}\,\ln \frac{\mu}{m_b}\, -
G(s_c,\bar u\,v)\, + \frac{1}{2}\,G(1,\bar u\,v)\,\right]\,
f_2(u,v)\, \biggl\}\,,\label{AQ10}
\end{eqnarray}
with the subscript $Q_{i}$ denoting the contraction of $Q_{i}$
operator in the weak  vertex, and
\begin{eqnarray}
f_1(u,v) &=& \Phi_{M_2}(u)\,\Phi_{M_1}(v)\,\left[\frac{2\,u + v -
3}{12\,( 1 - u ) \, {( 1 - v ) }^2} + \frac{3\,( 2\,u + v -3 )}
{4\,( 1 - u ) \, ( 1 - v ) \,v} \,\right]\,\nonumber \\
&+&
r_\chi^{M_1}\,\Phi_{M_2}(u)\,\Phi_{m_1}(v)\,\left[\frac{3\,( v
- 3)}{4\,( 1 - u ) \, ( 1 - v ) \,v} - \frac{1}{12\,( 1 - u ) \,
( 1 - v ) } \,\right]\,\nonumber\\
&+&
r_\chi^{M_1}\,r_\chi^{M_2}\,\Phi_{m_2}(u)\,\Phi_{m_1}(v)\,\left[\frac{2\,
u - 1}{12\,( 1 - u ) \,( 1 - v )} + \frac{3\,( 2\,u + v - 2\,u\,v
- 3)}
{4\,( 1 - u ) \,( 1 - v ) \,v} \right]\,\nonumber\\
&+&
r_\chi^{M_2}\,\Phi_{m_2}(u)\,\Phi_{M_1}(v)\,\left[\frac{v - 3}
{12\,{( 1 - v ) }^2} + \frac{3\,( v - 3 ) }{4\,( 1 - v ) \,v} \,\right]\,,\\
f_2(u,v) &=& \Phi_{M_2}(u)\,\Phi_{M_1}(v)\, \frac{1}{6\,{( 1 - v )
}^2}\, - r_\chi^{M_1}\,\Phi_{M_2}(u)\,\Phi_{m_1}(v)\,
\frac{1}{6\,u\,( 1 - v ) } \nonumber\\
&&+
r_\chi^{M_1}\,r_\chi^{M_2}\,\Phi_{m_2}(u)\,\Phi_{m_1}(v)\,\frac{1}{6\,(
1 - v )}\, +
r_\chi^{M_2}\,\Phi_{m_2}(u)\,\Phi_{M_1}(v)\,\frac{1}{6\,{( 1 - v ) }^2}\,\,,\\
f_1(u,v,m_q) & = &
\Phi_{M_2}(u)\,\Phi_{M_1}(v)\,\left[
\frac{3\,\Delta i_{2}}{8\,( 1 - u ) \,( 1 - v)} + \frac{3\,\Delta
i_{3}}{8\,( 1 - u ) \,v} + \frac{7\,\Delta i_{6}}{24\,( 1 - u ) \,
v}\,\right.\,\nonumber\\
&&\left.\, + \frac{3\,\Delta i_{8}}{8\,( 1 - v )\,v} +
\frac{7\,\Delta i_{23}}{24\,( 1 - v ) \,v} + \frac{7\,( 1 - u + v)
\,\Delta i_{5}}{24\,( 1 - u ) \,( 1 - v ) \,v}\,\right]\,\nonumber \\
&-&
r_\chi^{M_1}\,\Phi_{M_2}(u)\,\Phi_{m_1}(v)\,\left[
\frac{7\,\Delta i_{5}}{12\,( 1 - u ) \,( 1 - v ) } +
\frac{3\,(\Delta i_{2} - \Delta i_{8} + \Delta i_{17})}{8\,( 1 - u
)\, ( 1 - v ) } \,\right.\,\nonumber\\
&& \left. + \frac{7\,(\Delta i_{6} + \Delta i_{26})}{24\,( 1 - u
)\, v} + \frac{3\,(\Delta i_{3}+2\,\Delta i_{12}+\Delta
i_{21})}{8\,( 1 - u ) \,v} \,\right]\,\nonumber\\
&+&
r_\chi^{M_1}\,r_\chi^{M_2}\,\Phi_{m_2}(u)\,\Phi_{m_1}(v)\,\left[
\frac{7\,\Delta i_{5}}{12\,( 1 - v ) } + \frac{3\,(\Delta
i_{2}+\Delta i_{8}-\Delta i_{12}+\Delta i_{17})}{8\,v}\, \right. \,\nonumber\\
&& \left.- \frac{7\,u\,\Delta i_{23}}{12\,( 1 - u ) \,( 1 - v )} +
\frac{3\,(\Delta i_{2}+\Delta i_{8}+\Delta i_{17})}{8\,( 1 - u )
\,( 1 - v ) } + \frac{3\,u\,(\Delta i_{3}+\Delta i_{21})}{8\,( 1 -
u ) \,v} \,\right.\,\nonumber\\
&& \left. + \frac{7\,(\Delta i_{6}+\Delta i_{26})}{24\,( 1 - u ) \,v}\,\right]
\,\nonumber\\
&-& r_\chi^{M_2}\,\Phi_{m_2}(u)\,\Phi_{M_1}(v)\,\left[
\frac{3\,(\Delta i_{2}-\Delta i_{8})}{8\,( 1 - v ) \,v} +
\frac{7\,(\Delta i_{23}+2\,\Delta i_{5})}{24\,( 1 - v )\,v}\,\right]\,,\label{f1}\\
f_{2,(3)}(u,v,m_q) &=&
\Phi_{M_2}(u)\,\Phi_{M_1}(v)\,\left[-\frac{( 3 - 2\,u - 2\,v )
}{2\,( 1 - u ) \,( 1 - v ) \,v}+ \frac{3\,\Delta i_2}{8\,( 1 - u )
\,( 1 - v)} + \frac{3\,\Delta i_3}{8\,( 1 - u ) \,v}\,\right.\,\nonumber\\
&&\left.\pm \frac{7\,\Delta i_6}{24\,( 1 - u ) \,v}\,+
\frac{3\,\Delta i_8}{8\,( 1 - v )\,v} \pm \frac{7\,\Delta
i_{23}}{24\,( 1 - v ) \,v}\pm \frac{7\,( 1 - u + v ) \,\Delta
i_5}{24\,( 1 - u ) \,(1 - v ) \,v}\,\,\right]\,\nonumber \\
&-&
r_\chi^{M_1}\,\Phi_{M_2}(u)\,\Phi_{m_1}(v)\,\left[\frac{3}{2\,( 1
- u ) \,( 1 - v ) \,v}+ \frac{3\,(\Delta i_{3}+2\,\Delta
i_{12}+\Delta i_{21})}{8\,( 1 - u ) \,v}\,\right.\,\nonumber\\
&& \left. \pm \frac{7\,(\Delta i_{6}+\Delta i_{26})}{24\,( 1 - u )
\,v} \pm \frac{7\,\Delta i_5}{12\,( 1 - u )\,( 1 - v) } +
\frac{3\,(\Delta i_2 - \Delta i_8 + \Delta
i_{17})}{8\,( 1 - u )\,( 1 - v ) } \,\right]\,\nonumber\\
&+&
r_\chi^{M_1}\,r_\chi^{M_2}\,\Phi_{m_2}(u)\,\Phi_{m_1}(v)\,\left[-
\frac{3 - 2\,u - 2\,v + 2\,u\,v}{2\,( 1 - u ) \,( 1 - v ) \,v} \mp
\frac{7\,u\,\Delta i_{23}}{12\,( 1 - u ) \,( 1 - v ) } \,\right.\,\nonumber\\
&& \left. \pm \frac{7\,\Delta i_{5}}{12\,( 1 - v ) } +
\frac{3\,u\,(\Delta i_{3}+\Delta i_{21})}{8\,( 1 - u ) \,v}\, \pm
\frac{7\,(\Delta i_{6}+\Delta i_{26})}{24\,( 1 - u ) \,v}\,\right.\,\nonumber\\
&& \left. + \frac{3\,(\Delta i_{2}+\Delta i_{8}+\Delta
i_{17})}{8\,( 1 - u ) \,( 1 - v ) } \, + \frac{3\,(\Delta
i_{2}+\Delta i_{8}-\Delta i_{12}+\Delta i_{17})}{8\,v} \,\right]\,\nonumber\\
&-& r_\chi^{M_2}\,\Phi_{m_2}(u)\,\Phi_{M_1}(v)\,\left[
\frac{3}{2\,( 1 - v ) \,v} + \frac{3\,(\Delta i_{2}-\Delta
i_8)}{8\,( 1 - v ) \,v} \pm \frac{7\,(\Delta i_{23}+2\,\Delta
i_{5})}{24\,( 1 - v ) \,v}\,\right]\,,\label{f2}
\end{eqnarray}
when $M_1$ is a pseudoscalar and $M_2$ a vector meson. For the
opposite case, i.e., $M_1$ is a vector and $M_2$ a pseudoscalar
meson, one needs only change the signs of the last two terms in
the functions $f_i$ defined above. At this stage, the $\Delta_i$
functions appearing in Eqs.~(\ref{f1}) and (\ref{f2}) are the ones
that have been performed the Feynman parameter integrals, whose
explicit expressions can be found in Appendix B of
Refs.~\cite{Li:2005wx,yang:phiXs}. For convenience, we also list
them in Appendix A.

With the individual operator contribution given above, the total
contributions of the penguin contractions of spectator-scattering
amplitudes can be written as
\begin{eqnarray}\label{am3}
{\cal A}^{\prime}(B\to PV) &=& \frac{G_F}{\sqrt 2}\,\left[
\sum_{p=u,c} \! C_1\,{\cal A}_{Q_{1}} + (C_3\,-
\frac{1}{2}\,C_9)\,{\cal A}_{Q_{3}} +C_4\,{\cal A}_{Q_{4}} + C_6\,
{\cal A}_{Q_{6}}\,\right. \, \nonumber\\
&& \qquad\ \left.\, + C_8 \,{\cal A}_{Q_{8}} + C_{10}\,{\cal
A}_{Q_{10}} + {C_{8g}^{\rm eff}\,\cal A}_{Q_{8g}} \,\right]\,,
\end{eqnarray}
where the superscript `$\prime$' indicates the one to be
distinguished from the next-to-leading order results given by
Eqs.~(\ref{am1}) and~(\ref{am2}). The total decay amplitude is
then given as
\begin{equation}\label{amp}
{\langle}PV {\vert} {\cal H}_{\rm eff} {\vert}B{\rangle} =
 {\cal A}(B\to PV) + {\cal A}^{ann}(B\to PV) + {\cal A}^{\prime}(B\to
 PV)\,.
\end{equation}

\section{Numerical results and discussions}\label{sec3}

With the theoretical expressions given above and the input
parameters collected in Appendix B, we can now evaluate the
branching ratios and $CP$-violating asymmetries for two-body
hadronic charmless $B\to PV$ decays, with $P=(\pi, K)$ and
$V=(\rho, \omega, K^\ast, \phi)$. For each quantity, we first give
the results at next-to-leading order in $\alpha_s$, and then
take into account the higher order penguin contractions of
spectator-scattering amplitudes induced by the $b\to D g^{\ast}
g^{\ast}$ transitions. The combined contributions of these two
pieces, denoted by ${\cal O}(\alpha_s+\alpha_s^2)$, are then given
in the last. For comparison, results based on the NF approximation
are also presented. All the experimental data are taken from the
home page of the Heavy Flavor Averaging Group~(HFAG)~\cite{HFAG}.

In order to show the renormalization scale dependence of the
branching ratios and $CP$ asymmetries, we give results of two
cases for each decay mode with the first one evaluated at the
scale $\mu=m_b$, while the second at the scale $\mu=m_b/2$. In
addition, our calculations depend on many input parameters, which
cause quite large theoretical uncertainties. We will consider the
main theoretical uncertainties arising from the strange-quark
mass~(with the ratio $m_q/m_s$ fixed, all chiral enhancement
factors $r_{\chi}^P$ depend on this mass), CKM matrix elements,
form factors, and the first inverse moment of the $B$-meson
distribution amplitude $\lambda_B$.~\footnote{Since the
theoretical uncertainties coming from the weak annihilation and
twist-3 hard-spectator interaction contributions~(parameterized by
the quantities $X_H$ and $X_A$) are already known to be  quite
large\cite{bbns3}, we do not consider these uncertainties here and
simply use the default values given by
$X_{H(A)}=\log{m_B/\lambda_h}$ as specified in Appendix B.
Uncertainties coming from the other input parameters are generally
small and have been neglected.}

\subsection{Numerical analysis of penguin contractions of spectator-scattering
amplitudes}\label{sec3.1}

Before presenting numerical results for branching ratios and $CP$
asymmetries, we would discuss the relative strength of each
Feynman diagram shown in Figs.~\ref{Q8gfig} and \ref{penguinfig}.
For convenience, we denote the decay modes with the pseudoscalar
meson picking up the spectator quark by $B\to PV$, while for  the
vector meson picking up the spectator quark by $B\to VP$.

Firstly, we study the relative strength of the three Feynman
diagrams shown in Fig.~\ref{Q8gfig}. Since contributions of these
diagrams are all proportional to
\begin{equation}\label{factor1}
S_{1}=-i\,\frac{\alpha_s^2\, f_B \,f_{M_1}\,
f_{M_2}}{N_c^3}\,\lambda_t^{(\prime)}\,\int_0^1\!
d{\xi}\,\frac{\Phi_1^B (\xi)}{\xi}\,,
\end{equation}
we have factorized $S_{1}$ out off  the numerical results shown in
Table~\ref{tab:Q8g}.

\begin{table}[t]
\centerline{\parbox{16cm} {\caption{\label{tab:Q8g} Numerical
results of each Feynman diagram shown in Fig.~\ref{Q8gfig} with
the asymptotic forms of the meson LCDAs. Terms involving the
twist-three LCDAs are given in unit of the factor $r_{\chi}^{M}$
defined by Eq.~(\ref{chiral factor}). The subscripts $M_{1}=P(V)$
and $M_{2}=V(P) $ for the $B\to PV$ (  $B\to VP $ ) rows. The same
for $m_{1,2}$. }}} \vspace{0.1cm}
\begin{center}
\doublerulesep 0.8pt \tabcolsep 0.1in
\begin{tabular}{lcrrrrr}\hline\hline
 {}&{Decay mode}&$\Phi_{M_2}\Phi_{M_1}$ &
 $\Phi_{M_2}\Phi_{m_1}$ &
 $\Phi_{m_2}\Phi_{M_1}$ &
 $\Phi_{m_2}\Phi_{m_1}$&\\
\hline
 &$B\to PV$&$-67.50$&$-125.76$&$-9.64$&$-18.94$\\
 \raisebox{2.3ex}[0pt]{Fig.~\ref{Q8gfig}(a)}
 &$B\to VP$&$-67.50$&$4.82$&$34.71$&$-3.79$\\
\hline
 &$B\to PV$&$-1.50$&$-3.54$&$-1.07$&$-0.42$\\
 \raisebox{2.3ex}[0pt]{Figs.~\ref{Q8gfig}(b+c)}
 &$B\to VP$&$-1.50$&$-1.61$&$1.86$&$0.42$\\
\hline\hline
\end{tabular}
\end{center}
\end{table}

From the numerical results for the dipole operator $Q_{8g}$
contractions given in Table~\ref{tab:Q8g}, we can see that the
main contributions come from Fig.~\ref{Q8gfig}(a), and the other
ones play only a minor role. It is also noted that these
amplitudes  do not have strong phases.

To analyze strong phase sources and the relative strength of the
individual Feynman diagram shown in Fig.~\ref{penguinfig}, we
study the $Q_1^c$ contraction in the weak vertex. The contribution
of each Feynman diagram is proportional to
\begin{equation}\label{factor2}
S_{2}=i\,\frac{\alpha_s^2\, f_B \,f_{M_1}\,
f_{M_2}}{N_c^3}\,\lambda_c^{(\prime)}\,\int_0^1\!
d{\xi}\,\frac{\Phi_1^B (\xi)}{\xi}\,,
\end{equation}
which has also been factorized out. The numerical results given in
Table~\ref{tab:Q1c} are independent of $S_{2}$.

\begin{table}[t]
\centerline{\parbox{16cm} {\caption{\label{tab:Q1c} Numerical
results of each Feynman diagram shown in Fig.~\ref{penguinfig}. Others are the same as  Table~\ref{tab:Q8g}. }}} \vspace{0.1cm}
\begin{center}
\doublerulesep 0.8pt \tabcolsep 0.1in
\begin{tabular}{lcccccc}\hline\hline
 {}&modes &$\Phi_{M_2}\Phi_{M_1}$ &
 $\Phi_{M_2}\Phi_{m_1}$ &
 $\Phi_{m_2}\Phi_{M_1}$ &
 $\Phi_{m_2}\Phi_{m_1}$&\\
\hline
 &PV&$-1.39-12.65\,i$&$0.17-14.10\,i$&$-0.15+15.38\,i $&$0.12+13.51\,i$\\
 \raisebox{2.3ex}[0pt]{Fig.~\ref{penguinfig}(a)}
 &VP&$-1.39-12.65\,i$&$-0.02+1.28\,i$&$-0.12+11.11\,i$&$-0.01-0.44\,i$\\
\hline
 &PV&$-0.01-1.05\,i$&$-0.12-1.21\,i$&$-0.62+0.81\,i$&$-0.18-0.11\,i$\\
 \raisebox{2.3ex}[0pt]{Figs.~\ref{penguinfig}(b+c)}
 &VP&$-0.01-1.05\,i$&$-0.39-1.25\,i$&$-0.08+0.78\,i$&$-0.10-0.19\,i$\\
 \hline
 &PV&$-9.03+14.94\,i$&$19.19+28.30\,i$&$4.32-21.29\,i$&$10.82-15.69\,i$\\
 \raisebox{2.3ex}[0pt]{Figs.~\ref{penguinfig}(d+e)}
 &VP&$-9.03+14.94\,i$&$14.26+9.04\,i$&$0.83-16.78\,i$&$-0.39-3.46\,i$\\
\hline\hline
\end{tabular}
\end{center}
\end{table}

From the numerical results given in Table~\ref{tab:Q1c}, we have
the following observations: (i) contributions of
Figs.~\ref{penguinfig}(b) and \ref{penguinfig}(c) are generally
much smaller than those of the other three ones, and the main
contributions come from the diagrams Figs.~\ref{penguinfig}(d) and
\ref{penguinfig}(e); (ii) although each term labelled by the meson
LCDAs in each Feynman diagram has a large imaginary part, and
hence a large strong phase, the total strong phase of each Feynman
diagram is small due to cancellations among the four terms. (iii)
for each term labelled by the same LCDAs, there also exist
cancellations between the contributions of the diagrams
Fig.~\ref{penguinfig}(a) and Figs.~\ref{penguinfig}(d+e).

Thus the total strong phase is found to be quite small after
summing all the five diagrams shown in Fig.~\ref{penguinfig}.
Moreover, the cancellation does not depend on the parameters in
$S_{2}$.

\subsection{Branching ratios of $B\to PV$ decays}\label{sec3.2}

In the following discussions, we classify the two-body hadronic
charmless $B\to PV$ decays into two categories: the
strange-conserving~($\Delta S=0$) and the
strange-changing~($\Delta S=1$) processes. The higher order
penguin contractions of spectator-scattering amplitudes are
expected to have more significant impacts on the $\Delta S=1$
processes than on the $\Delta S=0$ ones, due to the CKM factor
suppressions in the latter.

Numerical results of the $CP$-averaged branching ratios for these
decays are collected in Tables~\ref{tab:br1}, \ref{tab:br2}, and
\ref{tab:br3}, where the theoretical error bars are due to the
uncertainties of the input parameters CKM elements, quark masses,
transition form factors, and $\lambda_{B}$ as collected in
Appendix B. Generally, the theoretical uncertainties are quite
large, which are larger than $\mathcal{O}(\alpha_{s}^{2})$
corrections for tree-dominated decay modes, but comparable for
strong penguin-dominated decay modes. For most decay modes, the
$\alpha_{s}^{2}$ corrections reduce the renormalization scale
dependence of the theoretical predictions.

\begin{table}[t]
\centerline{\parbox{16cm} {\caption{\label{tab:br1} $CP$-averaged
branching ratios~(in units of $10^{-6}$) of tree-dominated $B\to
PV$ decays with $\Delta S=0$. $\bar{{\cal B}}^f$ and $\bar{{\cal
B}}^{f+a}$ denote the results without and with the annihilation
contributions, respectively. Results in columns ${\cal
O}(\alpha_s+\alpha_s^2)$ are the ones with the higher order
penguin contraction contributions included. For each decay mode,
the first row is evaluated at the scale $\mu=m_b$, while the
second one at the scale $\mu=m_b/2$. The theoretical errors
correspond to the uncertainties of the input parameters collected
in Appendix B. The NF results are also shown for comparison.}}}
\vspace{0.1cm}
\begin{center}
\doublerulesep 0.8pt \tabcolsep 0.1in
\begin{tabular}{lccccccc}\hline\hline
 \multicolumn{2}{c@{\hspace{-9cm}}}{$\bar{\cal B}^f$} &
 \multicolumn{2}{c@{\hspace{-9cm}}}{$\bar{\cal B}^{f+a}$} \\
 \cline{3-4} \cline{5-6}\raisebox{4.3ex}[0pt]{Decay mode}
 &\raisebox{4.3ex}[0pt]{NF}& ${\cal O}(\alpha_s)$
 & ${\cal O}(\alpha_s+\alpha_s^2)$ & ${\cal O}(\alpha_s)$
 & ${\cal O}(\alpha_s+\alpha_s^2)$&\raisebox{4.3ex}[0pt]{EXP.}\\
\hline
 $B^{-} \to \pi^{-} \rho^{0}$
 &$8.76^{+3.56}_{-2.93}$&$8.15^{+3.69}_{-2.86}$&$8.02^{+3.77}_{-2.80}$
 &$8.13^{+3.53}_{-2.63}$&$8.01^{+3.73}_{-2.58}$&$8.7^{+1.0}_{-1.1}$\\
 &$7.52^{+3.36}_{-2.45}$&$7.45^{+3.42}_{-2.57}$&$7.36^{+3.71}_{-2.67}$
 &$7.44^{+3.25}_{-2.59}$&$7.36^{+3.60}_{-2.46}$&\\
 $B^{-} \to \pi^0 \rho^{-}$
 &$13.91^{+6.21}_{-4.87}$&$13.05^{+6.32}_{-4.53}$&$13.31^{+6.06}_{-4.76}$
 &$13.22^{+5.94}_{-4.80}$&$13.48^{+6.79}_{-5.05}$&$10.8^{+1.4}_{-1.5}$\\
 &$13.08^{+6.21}_{-4.54}$&$12.82^{+6.32}_{-4.86}$&$13.01^{+6.81}_{-5.16}$
 &$13.00^{+5.99}_{-4.94}$&$13.20^{+6.12}_{-4.89}$&\\
 $\overline{B}^0 \to \pi^{+} \rho^{-}$
 &$19.78^{+9.88}_{-7.28}$&$19.37^{+9.25}_{-7.62}$&$19.73^{+10.46}_{-7.28}$
 &$20.34^{+10.20}_{-7.95}$&$20.72^{+9.94}_{-7.85}$& $13.9^{+2.2}_{-2.1}$\\
 &$20.82^{+10.64}_{-7.83}$&$20.22^{+11.10}_{-8.11}$&$20.48^{+11.71}_{-7.65}$
 &$21.25^{+11.03}_{-8.26}$&$21.52^{+10.22}_{-7.86}$&\\
 $\overline{B}^0 \to \pi^{-} \rho^{+}$
 &$10.72^{+4.61}_{-3.68}$&$10.51^{+4.69}_{-3.55}$&$10.47^{+4.60}_{-3.49}$
 &$11.15^{+4.71}_{-3.82}$&$11.11^{+4.99}_{-3.75}$& $10.1^{+2.1}_{-1.9}$\\
 &$11.18^{+5.08}_{-3.74}$&$10.90^{+4.71}_{-3.89}$&$10.86^{+4.87}_{-3.92}$
 &$11.57^{+5.23}_{-4.02}$&$11.52^{+4.99}_{-3.90}$&\\
 $\overline{B}^0 \to \pi^{\pm} \rho^{\mp}$
 &$30.50^{+13.65}_{-10.39}$&$29.88^{+13.22}_{-10.18}$&$30.20^{+13.85}_{-10.52}$
 &$31.49^{+13.04}_{-10.64}$&$31.83^{+13.82}_{-11.48}$& $24.0\pm 2.5$\\
 &$32.00^{+14.58}_{-11.12}$&$31.12^{+14.60}_{-10.56}$&$31.34^{+13.82}_{-11.58}$
 &$32.82^{+14.96}_{-11.82}$&$33.04^{+16.32}_{-11.01}$&\\
 $\overline{B}^0 \to \pi^0 \rho^{0}$
 &$0.47^{+0.20}_{-0.15}$&$0.40^{+0.35}_{-0.18}$&$0.39^{+0.33}_{-0.15}$
 &$0.30^{+0.29}_{-0.13}$&$0.30^{+0.27}_{-0.13}$& $1.83^{+0.56}_{-0.55}$\\
 &$0.13^{+0.06}_{-0.04}$&$0.29^{+0.23}_{-0.12}$&$0.29^{+0.24}_{-0.11}$
 &$0.22^{+0.19}_{-0.08}$&$0.23^{+0.20}_{-0.09}$&\\
 $B^{-} \to \pi^{-} \omega$
 &$7.87^{+3.61}_{-2.57}$&$7.36^{+3.50}_{-2.44}$&$7.47^{+3.80}_{-2.53}$
 &$7.10^{+3.43}_{-2.62}$&$7.21^{+3.21}_{-2.37}$&$6.6 \pm 0.6$\\
 &$6.96^{+2.94}_{-2.28}$&$6.84^{+3.08}_{-2.39}$&$6.90^{+3.38}_{-2.31}$
 &$6.54^{+2.89}_{-2.23}$&$6.60^{+3.29}_{-2.28}$&\\
 $\overline{B}^0 \to \pi^{0} \omega$
 &$0.01^{+0.03}_{-0.01}$&$0.02^{+0.03}_{-0.01}$&$0.02^{+0.03}_{-0.01}$
 &$0.005^{+0.015}_{-0.003}$&$0.004^{+0.014}_{-0.003}$&$<1.2$\\
 &$0.03^{+0.04}_{-0.02}$&$0.02^{+0.02}_{-0.01}$&$0.02^{+0.03}_{-0.01}$
 &$0.010^{+0.018}_{-0.007}$&$0.010^{+0.020}_{-0.007}$&\\
\hline\hline
\end{tabular}
\end{center}
\end{table}

\begin{table}[t]
\centerline{\parbox{16cm} {\caption{\label{tab:br2} $CP$-averaged
branching ratios~(in units of $10^{-6}$) of penguin-dominated~(the
upper six) and annihilation-dominated~(the last two) $B\to PV$
decays with $\Delta S=0$. The captions are the same as
Table~\ref{tab:br1}.}}} \vspace{0.1cm}
\begin{center}
\doublerulesep 0.8pt \tabcolsep 0.1in
\begin{tabular}{lccccccc}\hline\hline
 \multicolumn{2}{c@{\hspace{-9cm}}}{$\bar{\cal B}^f$} &
 \multicolumn{2}{c@{\hspace{-9cm}}}{$\bar{\cal B}^{f+a}$} \\
 \cline{3-4} \cline{5-6}\raisebox{4.3ex}[0pt]{Decay mode}
 &\raisebox{4.3ex}[0pt]{NF}& ${\cal O}(\alpha_s)$
 & ${\cal O}(\alpha_s+\alpha_s^2)$ & ${\cal O}(\alpha_s)$
 & ${\cal O}(\alpha_s+\alpha_s^2)$&\raisebox{4.3ex}[0pt]{EXP.}\\
\hline
 $B^{-} \to K^{-} K^{\ast 0}$
 &$0.15^{+0.07}_{-0.04}$&$0.18^{+0.08}_{-0.07}$&$0.28^{+0.14}_{-0.09}$
 &$0.23^{+0.11}_{-0.09}$&$0.34^{+0.16}_{-0.11}$&$<5.3$\\
 &$0.32^{+0.13}_{-0.11}$&$0.23^{+0.10}_{-0.08}$&$0.33^{+0.15}_{-0.10}$
 &$0.29^{+0.14}_{-0.10}$&$0.40^{+0.20}_{-0.13}$&\\
 $\overline{B}^0 \to \overline{K}^{0} K^{\ast 0}$
 &$0.14^{+0.06}_{-0.04}$&$0.16^{+0.09}_{-0.06}$&$0.26^{+0.12}_{-0.08}$
 &$0.20^{+0.10}_{-0.07}$&$0.31^{+0.15}_{-0.10}$& $...$\\
 &$0.29^{+0.14}_{-0.09}$&$0.22^{+0.10}_{-0.08}$&$0.31^{+0.15}_{-0.10}$
 &$0.26^{+0.10}_{-0.09}$&$0.36^{+0.16}_{-0.11}$&\\
 $B^{-} \to K^{0} K^{\ast -}$
 &$0.06^{+0.13}_{-0.04}$&$0.10^{+0.21}_{-0.07}$&$0.10^{+0.20}_{-0.07}$
 &$0.18^{+0.27}_{-0.10}$&$0.18^{+0.26}_{-0.10}$&$...$\\
 &$0.05^{+0.14}_{-0.04}$&$0.08^{+0.18}_{-0.06}$&$0.07^{+0.17}_{-0.05}$
 &$0.15^{+0.25}_{-0.09}$&$0.14^{+0.23}_{-0.08}$&\\
 $\overline{B}^0 \to K^{0} \overline{K}^{\ast 0}$
 &$0.06^{+0.12}_{-0.04}$&$0.09^{+0.19}_{-0.06}$&$0.09^{+0.18}_{-0.06}$
 &$0.18^{+0.26}_{-0.10}$&$0.17^{+0.27}_{-0.09}$& $...$\\
 &$0.04^{+0.14}_{-0.03}$&$0.07^{+0.16}_{-0.05}$&$0.06^{+0.15}_{-0.04}$
 &$0.15^{+0.25}_{-0.08}$&$0.14^{+0.24}_{-0.08}$&\\
 $B^{-} \to \pi^{-} \phi$
 &$\approx0.001$&$\approx0.008$&$...$&$...$&$...$&$<0.41$\\
 &$\approx0.001$&$\approx0.007$&$...$&$...$&$...$&\\
 $\overline{B}^0 \to \pi^{0} \phi$
 &$\approx0.0003$&$\approx0.004$&$...$&$...$&$...$&$<1.0$\\
 &$\approx0.0003$&$\approx0.003$&$...$&$...$&$...$&\\
 $\overline{B}^0 \to K^{\ast -} K^{+}$
 & $...$&$...$&$...$&$0.018^{+0.004}_{-0.004}$&$...$& $...$\\
 & $...$&$...$&$...$&$0.019^{+0.005}_{-0.004}$&$...$& $...$\\
 $\overline{B}^0 \to K^{-} K^{\ast +}$
 & $...$&$...$&$...$&$0.018^{+0.004}_{-0.004}$&$...$&\\
 & $...$&$...$&$...$&$0.019^{+0.005}_{-0.004}$&$...$&\\
\hline\hline
\end{tabular}
\end{center}
\end{table}

\begin{table}[htbp]
\centerline{\parbox{16cm} {\caption{\label{tab:br3} $CP$-averaged
branching ratios (in units of $10^{-6}$) of penguin-dominated
$B\to PV$ decays with $\Delta S=1$. The captions are the same as
Table~\ref{tab:br1}. }}} \vspace{0.1cm}
\begin{center}
\doublerulesep 0.8pt \tabcolsep 0.1in
\begin{tabular}{lccccccc}\hline\hline
 \multicolumn{2}{c@{\hspace{-9cm}}}{$\bar{\cal B}^f$} &
 \multicolumn{2}{c@{\hspace{-9cm}}}{$\bar{\cal B}^{f+a}$} \\
 \cline{3-4} \cline{5-6}\raisebox{4.3ex}[0pt]{Decay mode}
 &\raisebox{4.3ex}[0pt]{NF}& ${\cal O}(\alpha_s)$
 & ${\cal O}(\alpha_s+\alpha_s^2)$ & ${\cal O}(\alpha_s)$
 & ${\cal O}(\alpha_s+\alpha_s^2)$& \raisebox{4.3ex}[0pt]{EXP.}\\
\hline
 $B^{-} \to \pi^{-} \overline{K}^{\ast 0} $
 &$2.37^{+0.72}_{-0.64}$&$2.60^{+0.95}_{-0.88}$&$4.26^{+1.72}_{-1.21}$
 &$3.50^{+1.22}_{-1.04}$&$5.39^{+2.01}_{-1.44}$&$10.8 \pm 0.8$\\
 &$4.89^{+1.46}_{-1.28}$&$3.35^{+1.27}_{-1.13}$&$5.01^{+1.81}_{-1.41}$
 &$4.45^{+1.51}_{-1.36}$&$6.34^{+2.18}_{-1.70}$&\\
 $B^{-} \to \pi^0 K^{\ast -}$
 &$1.82^{+0.76}_{-0.54}$&$1.88^{+0.79}_{-0.56}$&$2.73^{+1.23}_{-0.81}$
 &$2.33^{+0.96}_{-0.69}$&$3.29^{+1.31}_{-0.89}$&$6.9 \pm 2.3$\\
 &$3.03^{+1.15}_{-0.88}$&$2.21^{+0.87}_{-0.74}$&$3.05^{+1.25}_{-0.89}$
 &$2.75^{+1.08}_{-0.79}$&$3.70^{+1.36}_{-1.01}$&\\
 $\overline{B}^0 \to \pi^{+} K^{\ast -} $
 &$1.84^{+0.90}_{-0.67}$&$1.92^{+0.89}_{-0.72}$&$3.04^{+1.64}_{-1.04}$
 &$2.47^{+1.08}_{-0.82}$&$3.78^{+1.84}_{-1.34}$& $11.7^{+1.5}_{-1.4}$\\
 &$3.40^{+1.49}_{-1.11}$&$2.32^{+1.12}_{-0.84}$&$3.43^{+1.67}_{-1.13}$
 &$2.99^{+1.31}_{-0.96}$&$4.30^{+2.09}_{-1.44}$&\\
 $\overline{B}^0 \to \pi^0 \overline{K}^{\ast 0}$
 &$0.49^{+0.27}_{-0.20}$&$0.53^{+0.35}_{-0.26}$&$1.08^{+0.77}_{-0.46}$
 &$0.80^{+0.42}_{-0.33}$&$1.45^{+0.86}_{-0.56}$& $1.7 \pm 0.8$\\
 &$1.24^{+0.56}_{-0.46}$&$0.73^{+0.50}_{-0.35}$&$1.28^{+0.73}_{-0.50}$
 &$1.07^{+0.56}_{-0.43}$&$1.72^{+0.91}_{-0.65}$&\\
 $B^{-} \to K^{-} \phi$
 & $3.71^{+1.18}_{-1.00}$&$2.73^{+1.33}_{-1.20}$&$5.06^{+2.01}_{-1.48}$
 &$4.04^{+1.58}_{-1.48}$&$6.77^{+2.78}_{-1.74}$&$9.03^{+0.65}_{-0.63}$\\
 &$10.17^{+3.21}_{-3.23}$&$3.90^{+1.93}_{-1.69}$&$6.32^{+2.07}_{-1.77}$
 &$5.59^{+2.23}_{-2.11}$&$8.42^{+2.67}_{-2.22}$&\\
 $\overline{B}^0 \to \overline{K}^{0} \phi$
 &$3.45^{+1.10}_{-0.93}$&$2.53^{+1.20}_{-1.11}$&$4.70^{+1.86}_{-1.37}$
 &$3.67^{+1.50}_{-1.37}$&$6.19^{+2.40}_{-1.69}$&$8.3^{+1.2}_{-1.0}$\\
 &$9.46^{+3.01}_{-2.59}$&$3.63^{+1.81}_{-1.61}$&$5.88^{+2.10}_{-1.67}$
 &$5.09^{+2.10}_{-1.87}$&$7.70^{+2.55}_{-2.14}$&\\
 $B^{-} \to \overline{K}^{0} \rho^{-} $
 &$1.05^{+2.12}_{-0.73}$&$1.74^{+3.09}_{-1.16}$&$1.65^{+3.10}_{-1.08}$
 &$3.18^{+4.42}_{-1.85}$&$3.05^{+3.94}_{-1.73}$&$<48$\\
 &$0.76^{+2.17}_{-0.63}$&$1.36^{+2.99}_{-0.97}$&$1.20^{+2.69}_{-0.86}$
 &$2.73^{+3.77}_{-1.58}$&$2.49^{+3.83}_{-1.47}$&\\
 $B^{-} \to K^{-} \rho^{0}$
 &$0.77^{+1.06}_{-0.35}$&$0.99^{+1.70}_{-0.59}$&$0.96^{+1.69}_{-0.56}$
 &$1.56^{+2.38}_{-0.95}$&$1.51^{+2.24}_{-0.95}$&$4.23^{+0.56}_{-0.57}$\\
 &$0.58^{+1.11}_{-0.26}$&$0.78^{+1.56}_{-0.43}$&$0.72^{+1.35}_{-0.36}$
 &$1.28^{+2.10}_{-0.78}$&$1.19^{+2.12}_{-0.70}$&\\
 $\overline{B}^0 \to K^{-} \rho^{+} $
 &$2.50^{+3.17}_{-1.36}$&$3.44^{+4.20}_{-1.91}$&$3.31^{+4.09}_{-1.81}$
 &$5.27^{+5.29}_{-2.67}$&$5.11^{+5.18}_{-2.55}$& $9.9^{+1.6}_{-1.5}$\\
 &$2.28^{+3.33}_{-1.33}$&$3.04^{+3.66}_{-1.69}$&$2.81^{+3.77}_{-1.54}$
 &$4.86^{+5.19}_{-2.42}$&$4.55^{+5.00}_{-2.32}$&\\
 $\overline{B}^0 \to \overline{K}^{0} \rho^0 $
 &$1.42^{+1.59}_{-0.72}$&$1.98^{+2.13}_{-1.03}$&$1.90^{+2.12}_{-0.97}$
 &$3.03^{+3.01}_{-1.35}$&$2.94^{+2.68}_{-1.39}$& $5.1 \pm 1.6$\\
 &$1.32^{+1.79}_{-0.76}$&$1.80^{+2.17}_{-0.94}$&$1.66^{+1.97}_{-0.95}$
 &$2.88^{+2.61}_{-1.35}$&$2.70^{+2.59}_{-1.27}$&\\
 $B^{-} \to K^{-} \omega$
 &$0.89^{+1.18}_{-0.48}$&$2.16^{+2.33}_{-1.12}$&$2.10^{+2.55}_{-1.11}$
 &$3.07^{+3.01}_{-1.49}$&$2.99^{+3.07}_{-1.44}$&$6.5 \pm 0.6$\\
 &$0.40^{+0.87}_{-0.13}$&$1.75^{+2.15}_{-0.97}$&$1.65^{+2.27}_{-0.94}$
 &$2.61^{+3.20}_{-1.42}$&$2.47^{+3.25}_{-1.29}$&\\
 $\overline{B}^0 \to \overline{K}^{0} \omega$
 &$0.17^{+0.66}_{-0.15}$&$1.03^{+1.74}_{-0.68}$&$0.99^{+1.67}_{-0.66}$
 &$1.78^{+2.45}_{-1.00}$&$1.72^{+2.26}_{-0.96}$&$4.7 \pm 0.6$\\
 &$0.03^{+0.29}_{-0.03}$&$0.76^{+1.49}_{-0.52}$&$0.69^{+1.45}_{-0.47}$
 &$1.43^{+2.16}_{-0.83}$&$1.33^{+2.09}_{-0.82}$&\\
\hline\hline
\end{tabular}
\end{center}
\end{table}

For $\Delta S=0$ decays, since the $b\to d$ penguin amplitudes are
suppressed by the CKM factor $\lambda_t^{\prime}$ compared to the
$b\to s$ penguin amplitudes, most of them are expected to be
dominated by the tree amplitudes, however with a few exceptions.
From the numerical results given in Tables~\ref{tab:br1} and
\ref{tab:br2}, we have the following general remarks:
\begin{itemize}

\item The decays $\overline{B}^0\to\pi^{\pm}\rho^{\mp}$ and
$B^{-}\to\pi^{0}\rho^{-}, \pi^{-}\rho^{0}, \pi^{-}\omega$. Our
results are generally consistent with the experimental data within
errors. Since these decay channels are dominated by the
color-allowed tree amplitudes, both the weak annihilation and the
higher order penguin contraction contributions are small. In
addition, the main theoretical errors come from the uncertainties
of the form factors and CKM matrix elements.

\item The decays $\overline{B}^0\to \pi^{0}\rho^{0}$ and
$\overline{B}^0\to \pi^{0}\omega$. Since these decay channels are
dominated by the color-suppressed tree amplitudes, their branching
ratios are predicted to be very small. The higher order penguin
contraction contributions are always much smaller than the weak
annihilation contributions. Besides the form factors and CKM
matrix elements, the spectator-scattering amplitudes also cause
sizable uncertainties to their $CP$-averaged branching ratios.

\item The decays $B^{-}\to K^{-} K^{*0}$ and
$\overline{B}^0\to\overline{K}^0 K^{*0}$. These decay channels are
dominated by the $b\to d$ penguin amplitudes, and the dominant
term is proportional to the coefficient $\alpha_4^p(PV)$. Since
$\alpha_4^c\approx \alpha_4^u$ and
$|\lambda_u^{\prime}|\approx|\lambda_c^{\prime}|$, large
interference effects between these two terms are expected and the
branching ratios of these decay modes have a strong dependence on
the weak phase angle $\gamma=arg(V_{ub}^{\ast})$. The higher order
penguin contraction contributions can provide about $60\%$
enhancements to their branching ratios, and are larger than the
weak annihilation contributions~(which also play an important role
in these decay channels). Since the higher order penguin
contraction contributions are all involved the quantity
$\lambda_B$, the main theoretical errors in the $CP$-averaged
branching ratios, besides the CKM matrix elements and form
factors, also originate from this quantity.

\item The decays $ B^{-}\to K^{0}K^{*-}$ and $\overline{B}^0\to
K^0\overline{K}^{*0}$. The dominant contribution to the decay
amplitudes is proportional to the coefficient $\alpha_4^p(VP)$,
where delicate cancellations between the vector and scalar penguin
contributions occur, their branching ratios are therefore
predicted to be relatively small. This also renders the weak
annihilation contributions potentially large. On the other hand,
since these decay channels belong to the category of $B\to VP$
decays, the higher order penguin contraction contributions are
predicted to be small. The theoretical errors in the $CP$-averaged
branching ratios of these decay channels are large, mainly due to
the variations of the strange-quark mass and $\lambda_B$.

\item The decays $ B^-\to \pi^{-}\phi $ and
$\overline{B}^0\to\pi^{0}\phi$. These two decay channels do not
receive the weak annihilation contributions and are electro-weak
penguin dominated processes. Due to the small coefficients
$\alpha_3^p(\pi\phi)$ and $\alpha_{3,ew}^p(\pi\phi)$, their
branching ratios are predicted to be quite small. From the
numerical results, we can see that large ``nonfactorizable"
contributions dominate these decays, while the theoretical
predictions are still quite lower than the experimental upper
bounds. The higher order penguin contraction contributions have
negligible impact on these decay channels.

\item The decays $\overline{B}^0\to K^{+}K^{*-}, K^{-}K^{*+}$.
These two decay channels are pure annihilation processes. Studying
on these decay modes may be helpful to learn more about the
strength of annihilation contributions and to provide some useful
information about final-state interactions.  The higher order
penguin contraction contributions have no impacts on these decay
channels.
\end{itemize}

For penguin-dominated $\Delta S=1$ decays, since the QCD penguin
coefficients $\alpha_{3,4}^p$ can receive large
``nonfactorizable'' contributions within the QCDF formalism, the
predicted branching ratios for these decay modes are usually quite
different from those obtained with the NF approximation. In
addition, the weak annihilation contributions to these decay
channels are quite sizable. From the numerical results given in
Table~\ref{tab:br3}, we have the following general remarks:
\begin{itemize}

\item The decays $B\to\pi K^{*}$ and $B\to K \phi$. With central
values of our input parameters, our results are still lower than
the experimental data. The dominant contribution to the decay
amplitudes is proportional to the coefficient $\alpha_4^p(PV)$.
The higher order penguin contraction contributions can give
enhancements to these branching ratios by about $40\%\sim90\%$,
and reduce the discrepancies between the theoretical predictions
and the experimental data. In addition, large interference effects
between the tree and penguin amplitudes in some decay channels,
such as $\overline{B}^0 \to \pi^{+} K^{\ast -}$ and $B^{-} \to
\pi^{0} K^{\ast -}$, are expected. It is thus possible to gain
some information on the weak angle $\gamma$ from these decay
channels. The main theoretical errors are due to the uncertainties
of the CKM matrix elements, form factors, and $\lambda_B$.

\item The decays $B\to K \rho$ and $B\to K \omega$. In their decay
amplitudes, the dominant term is proportional to the coefficient
$\alpha^p_4(VP)$. Because of the destructive interference between
the vector and the scalar penguin contributions, the coefficient
$\alpha_4^p(VP)$ is reduced to a large extent, making the
branching ratios of these decay modes much smaller than those of
the corresponding $B\to PP$ counterparts. It also makes the
sub-leading terms, for example, the weak annihilation
contributions, very important to account for the experimental
data. Since these decay channels also belong to the category of
$B\to VP$ decays, the higher order penguin contraction
contributions are quite small, and tend to decrease the NLO
results. The main theoretical errors are due to the uncertainties
of the strange-quark mass and form factors.
\end{itemize}

From the above discussions, we can see that the higher order
penguin contractions of spectator-scattering amplitudes play an
important role in penguin-dominated $B\to PV$ decays, while for
tree-dominated $B\to PV$ decays, their effects are generally quite
small. In particular, for decay modes dominated by the coefficient
$\alpha_4^p(PV)$, these higher order penguin contraction
contributions can increase the branching ratios by about
$40\%\sim90\%$, while for those dominated by the coefficient
$\alpha_4^p(VP)$, their contributions are also predicted to be
small and tend to decrease the branching ratios of these decay
modes. At present, all these predicted $CP$-averaged branching
ratios still suffer from large theoretical uncertainties.

\subsection{Direct $CP$-violating asymmetries of $B\to PV$ decays}
\label{sec3.3}

In this subsection, we will discuss the direct $CP$-violating
asymmetries. In particular, we will investigate the impact of the
higher order penguin contractions of spectator-scattering
amplitudes on this quantity.

\begin{table}[t]
\centerline{\parbox{16cm} {\caption{\label{tab:cp1} Direct
$CP$-violating asymmetries~(in units of $10^{-2}$) for two-body
hadronic charmless $B\to PV$ decays with $\Delta S=0$. Decay modes
with very small branching ratios are not considered. ${\cal
A}_{CP}^f$ and ${\cal A}_{CP}^{f+a}$ denote the results without
and with the annihilation contributions, respectively. The other
captions are the same as Table~\ref{tab:br1}.}}}
\begin{center}
\doublerulesep 0.8pt \tabcolsep 0.1in
\begin{tabular}{lcccccccc}\hline\hline
 \multicolumn{2}{c@{\hspace{-5cm}}}{${\cal A}_{CP}^f$} &
 \multicolumn{2}{c@{\hspace{-5cm}}}{${\cal A}_{CP}^{f+a}$} \\
 \cline{2-3} \cline{4-5}\raisebox{4.3ex}[0pt]{Decay mode}
 &${\cal O}(\alpha_s)$ & ${\cal O}(\alpha_s+\alpha_s^2)$
 &${\cal O}(\alpha_s)$ & ${\cal O}(\alpha_s+\alpha_s^2)$
 &\raisebox{4.3ex}[0pt]{EXP.}\\
\hline
 $B^{-} \to \pi^{-} \rho^{0}$
 &$3.25^{+1.98}_{-1.27}$&$5.26^{+3.62}_{-2.06}$&
 $3.62^{+2.29}_{-1.43}$&$5.64^{+3.58}_{-2.13}$&$-7^{+12}_{-13}$\\
 &$2.83^{+2.35}_{-1.33}$&$4.02^{+3.04}_{-1.58}$
 &$3.39^{+2.36}_{-1.53}$&$4.58^{+2.78}_{-1.78}$&\\
 $B^{-} \to \pi^0 \rho^{-}$
 &$-2.41^{+0.81}_{-1.61}$&$-3.69^{+1.39}_{-2.48}$
 &$-2.63^{+0.83}_{-1.63}$&$-3.88^{+1.37}_{-2.52}$&$1\pm 11$\\
 &$-1.74^{+0.68}_{-1.54}$&$-2.49^{+0.94}_{-1.84}$
 &$-2.03^{+0.76}_{-1.70}$&$-2.76^{+0.95}_{-1.87}$&\\
 $\overline{B}^0 \to \pi^{+} \rho^{-}$
 &$-1.05^{+0.12}_{-0.19}$&$-2.65^{+0.92}_{-1.85}$
 &$-1.03^{+0.12}_{-0.17}$&$-2.57^{+0.80}_{-1.82}$&$-15\pm 9$\\
 &$-0.68^{+0.08}_{-0.11}$&$-1.68^{+0.45}_{-1.03}$
 &$-0.65^{+0.07}_{-0.11}$&$-1.62^{+0.44}_{-0.89}$&\\
 $\overline{B}^0 \to \pi^{-} \rho^{+}$
 &$0.40^{+0.64}_{-0.37}$&$-0.03^{+0.64}_{-0.60}$
 &$0.31^{+0.58}_{-0.37}$&$-0.13^{+0.64}_{-0.53}$&$-47^{+13}_{-14}$\\
 &$-0.76^{+0.23}_{-0.27}$&$-1.36^{+0.41}_{-0.65}$
 &$-0.88^{+0.23}_{-0.29}$&$-1.49^{+0.40}_{-0.64}$&\\
 $\overline{B}^0 \to \pi^0 \rho^{0}$
 &$-5.64^{+9.80}_{-17.89}$&$5.92^{+10.14}_{-17.18}$
 &$-13.49^{+11.83}_{-20.61}$&$-0.22^{+12.35}_{-23.54}$&$-49^{+70}_{-83}$\\
 &$-4.42^{+19.18}_{-28.38}$&$10.58^{+18.83}_{-28.48}$
 &$-19.13^{+18.98}_{-32.25}$&$-1.68^{+21.36}_{-34.52}$&\\
 $B^{-} \to \pi^{-} \omega$
 &$-1.95^{+1.54}_{-2.03}$&$-4.49^{+1.68}_{-2.34}$
 &$-1.84^{+1.58}_{-2.09}$&$-4.45^{+1.66}_{-2.16}$&$-4 \pm 8$\\
 &$-4.46^{+2.11}_{-3.14}$&$-6.66^{+2.38}_{-3.37}$
 &$-4.36^{+2.10}_{-3.09}$&$-6.64^{+2.42}_{-3.21}$&\\
 $B^{-} \to K^{-} K^{\ast 0}$
 &$-36.28^{+5.04}_{-5.51}$&$-19.29^{+8.89}_{-6.15}$
 &$-31.08^{+4.37}_{-4.67}$&$-15.34^{+8.74}_{-6.47}$&$...$\\
 &$-42.06^{+5.68}_{-6.38}$&$-28.33^{+6.84}_{-5.54}$
 &$-36.92^{+5.40}_{-5.29}$&$-24.27^{+6.78}_{-5.82}$&\\
 $\overline{B}^0 \to \overline{K}^{0} K^{\ast 0}$
 &$-36.27^{+5.02}_{-5.66}$&$-19.29^{+8.34}_{-6.48}$
 &$-32.72^{+4.74}_{-4.82}$&$-17.56^{+7.65}_{-5.57}$& $...$\\
 &$-42.06^{+5.43}_{-6.50}$&$-28.33^{+6.91}_{-5.56}$
 &$-38.64^{+5.15}_{-5.46}$&$-26.25^{+6.10}_{-6.04}$&\\
 $B^{-} \to K^{0} K^{\ast -}$
 &$-12.64^{+4.49}_{-4.14}$&$-22.25^{+4.35}_{-7.40}$
 &$-9.41^{+5.03}_{-4.82}$&$-15.93^{+4.95}_{-4.54}$&$...$\\
 &$-2.96^{+8.53}_{-6.64}$&$-18.26^{+5.22}_{-9.82}$
 &$0.18^{+10.23}_{-7.16}$&$-9.17^{+8.89}_{-6.79}$&\\
 $\overline{B}^0 \to K^{0} \overline{K}^{\ast 0}$
 &$-12.64^{+4.60}_{-4.00}$&$-22.25^{+4.24}_{-8.09}$
 &$-9.25^{+4.55}_{-4.78}$&$-16.25^{+4.90}_{-4.25}$& $...$\\
 &$-2.96^{+8.64}_{-6.76}$&$-18.26^{+5.59}_{-8.60}$
 &$-1.76^{+6.45}_{-5.73}$&$-12.22^{+5.90}_{-6.34}$&\\
\hline\hline
\end{tabular}
\end{center}
\end{table}

\begin{table}[htbp]
\centerline{\parbox{16cm} {\caption{\label{tab:cp2} Direct
$CP$-violating asymmetries~(in units of $10^{-2}$) for two-body
hadronic charmless $B\to PV$ decays with $\Delta S=1$. The
captions are the same as Table~\ref{tab:cp1}.}}}
\begin{center}
\doublerulesep 0.8pt \tabcolsep 0.1in
\begin{tabular}{lcccccc}\hline\hline
 \multicolumn{2}{c@{\hspace{-5cm}}}{${\cal A}_{CP}^f$} &
 \multicolumn{2}{c@{\hspace{-5cm}}}{${\cal A}_{CP}^{f+a}$} \\
 \cline{2-3} \cline{4-5}\raisebox{4.3ex}[0pt]{Decay mode}
 &${\cal O}(\alpha_s)$ & ${\cal O}(\alpha_s+\alpha_s^2)$
 &${\cal O}(\alpha_s)$ & ${\cal O}(\alpha_s+\alpha_s^2)$
 &\raisebox{4.3ex}[0pt]{EXP.}\\
\hline
 $B^{-} \to \pi^{-} \overline{K}^{\ast 0} $
 &$1.49^{+0.23}_{-0.14}$&$0.76^{+0.26}_{-0.34}$
 &$1.22^{+0.14}_{-0.13}$&$0.57^{+0.26}_{-0.34}$&$-9.3\pm 6.0$\\
 &$1.77^{+0.24}_{-0.17}$&$1.14^{+0.21}_{-0.24}$
 &$1.47^{+0.17}_{-0.15}$&$0.93^{+0.21}_{-0.26}$&\\
 $B^{-} \to \pi^0 K^{\ast -}$
 &$14.03^{+2.88}_{-2.44}$&$18.21^{+5.43}_{-4.15}$
 &$11.98^{+2.46}_{-2.10}$&$15.48^{+4.69}_{-3.59}$&$4\pm 29$\\
 &$13.09^{+3.48}_{-2.66}$&$14.85^{+3.47}_{-3.07}$
 &$11.27^{+2.74}_{-2.40}$&$12.72^{+2.74}_{-2.43}$&\\
 $\overline{B}^0 \to \pi^{+} K^{\ast -} $
 &$9.14^{+1.51}_{-1.34}$&$17.18^{+6.39}_{-4.76}$
 &$7.11^{+1.31}_{-1.24}$&$13.75^{+5.50}_{-4.06}$& $-5\pm 14$\\
 &$3.89^{+0.65}_{-0.59}$&$9.16^{+2.87}_{-2.09}$
 &$2.86^{+0.52}_{-0.49}$&$7.16^{+1.93}_{-1.46}$&\\
 $\overline{B}^0 \to \pi^0 \overline{K}^{\ast 0}$
 &$-11.58^{+4.15}_{-8.58}$&$-9.94^{+3.14}_{-4.69}$
 &$-9.20^{+2.79}_{-5.00}$&$-8.34^{+2.64}_{-3.77}$& $-1^{+27}_{-26}$\\
 &$-12.14^{+4.04}_{-7.46}$&$-10.06^{+3.09}_{-4.31}$
 &$-9.97^{+3.36}_{-4.79}$&$-8.60^{+2.47}_{-3.68}$&\\
 $B^{-} \to K^{-} \phi$
 &$2.08^{+0.53}_{-0.27}$&$1.07^{+0.32}_{-0.37}$
 &$1.61^{+0.23}_{-0.18}$&$0.78^{+0.30}_{-0.39}$&$3.7\pm 5.0$\\
 &$2.33^{+0.56}_{-0.31}$&$1.49^{+0.21}_{-0.22}$
 &$1.84^{+0.27}_{-0.20}$&$1.17^{+0.23}_{-0.23}$&\\
 $\overline{B}^0 \to \overline{K}^{0} \phi$
 &$2.08^{+0.50}_{-0.27}$&$1.07^{+0.33}_{-0.39}$
 &$1.72^{+0.27}_{-0.19}$&$0.92^{+0.25}_{-0.39}$&$9\pm 14$\\
 &$2.33^{+0.58}_{-0.29}$&$1.49^{+0.20}_{-0.23}$
 &$1.96^{+0.33}_{-0.21}$&$1.30^{+0.19}_{-0.25}$&\\
 $B^{-} \to \overline{K}^{0} \rho^{-} $
 &$0.49^{+0.14}_{-0.17}$&$0.93^{+0.34}_{-0.15}$
 &$0.37^{+0.17}_{-0.21}$&$0.67^{+0.20}_{-0.18}$&$...$\\
 &$0.11^{+0.25}_{-0.32}$&$0.80^{+0.41}_{-0.20}$
 &$-0.02^{+0.29}_{-0.40}$&$0.41^{+0.28}_{-0.36}$&\\
 $B^{-} \to K^{-} \rho^{0}$
 &$-7.99^{+11.58}_{-5.17}$&$-3.62^{+17.39}_{-6.87}$
 &$-7.32^{+4.63}_{-3.55}$&$-4.55^{+8.50}_{-4.26}$&$31^{+12}_{-11}$\\
 &$5.88^{+27.17}_{-10.73}$&$15.31^{+33.85}_{-16.23}$
 &$-0.38^{+13.62}_{-5.64}$&$5.17^{+23.92}_{-8.71}$&\\
 $\overline{B}^0 \to K^{-} \rho^{+} $
 &$-1.76^{+1.64}_{-0.87}$&$0.24^{+4.86}_{-1.84}$
 &$-0.91^{+1.21}_{-0.85}$&$0.46^{+2.99}_{-1.42}$&$17^{+15}_{-16}$\\
 &$4.12^{+4.76}_{-2.50}$&$7.89^{+10.46}_{-4.88}$
 &$3.02^{+3.08}_{-1.67}$&$5.44^{+6.24}_{-3.07}$&\\
 $\overline{B}^0 \to \overline{K}^{0} \rho^0 $
 &$9.58^{+3.69}_{-3.24}$&$9.73^{+3.86}_{-3.29}$
 &$7.65^{+2.85}_{-2.30}$&$7.78^{+2.67}_{-2.45}$& $...$\\
 &$12.36^{+5.78}_{-4.30}$&$12.91^{+5.89}_{-4.81}$
 &$9.81^{+3.63}_{-3.16}$&$10.23^{+4.29}_{-3.46}$&\\
 $B^{-} \to K^{-} \omega$
 &$-4.71^{+2.93}_{-2.41}$&$-2.85^{+4.26}_{-3.31}$
 &$-4.35^{+2.05}_{-1.93}$&$-3.04^{+2.92}_{-2.35}$&$2\pm 7$\\
 &$4.75^{+13.57}_{-5.57}$&$8.69^{+16.81}_{-7.30}$
 &$1.39^{+6.29}_{-3.35}$&$3.94^{+9.10}_{-4.60}$&\\
 $\overline{B}^0 \to \overline{K}^{0} \omega$
 &$-9.65^{+4.10}_{-5.65}$&$-8.90^{+3.91}_{-5.41}$
 &$-7.61^{+2.96}_{-4.62}$&$-7.13^{+2.69}_{-3.99}$&$44\pm 23$\\
 &$-12.85^{+5.95}_{-6.22}$&$-11.61^{+5.50}_{-5.40}$
 &$-10.55^{+4.54}_{-7.83}$&$-9.94^{+4.27}_{-6.60}$&\\
\hline\hline
\end{tabular}
\end{center}
\end{table}

We adopt the convention for direct $CP$ asymmetries
\begin{equation}\label{acpdef}
{\cal A}_{CP}\equiv \frac{{\cal B}(\overline B^0\to\bar f) - {\cal
B}(B^0\to f)} {{\cal B}(\overline B^0\to\bar f) + {\cal B}(B^0\to
f)} \,.
\end{equation}
Our numerical results for the direct $CP$-violating asymmetries
are listed in Tables~\ref{tab:cp1} and \ref{tab:cp2}. Since the
strong phases are suppressed by $\alpha_s$ and/or $\Lambda_{\rm
QCD}/m_b$ within the QCDF formalism, the direct $CP$-violating
asymmetries for most $B\to PV$ decays are predicted to be
typically small. This is particularly true for decay modes
dominated by the tree coefficient $\alpha_1$, for example, the
decay $\overline{B}^0 \to \pi^{-} \rho^{+}$. However, for $b\to d$
penguin dominated $B\to K {\bar K}^\ast$ decays, the penguin
amplitudes generated by the internal u-quark loop  and c-quark
loop are proportional to the comparable CKM elements
$V_{ub}^{*}V_{ud}$ and $V_{cb}^{*}V_{cd}$, respectively, large
direct $CP$-violating asymmetries for these decay channels are
predicted. In addition, due to large interference effects between
the tree and penguin amplitudes, the direct $CP$-violating
asymmetry of $B^-\to \pi^0 K^{\ast -}$ decay is also predicted to
be large.

Due to cancellations among the strong phases associated with the
individual Feynman diagram in Fig.~\ref{penguinfig} as discussed
in Sec.~\ref{sec3.1}, the higher order penguin contraction
contributions to the direct $CP$-violating asymmetries for most
$B\to PV$ decays are predicted to be small, however with a few
exceptions.   From  Table~\ref{tab:cp1}, we can see that both the
higher order penguin contraction and the weak annihilation
contributions have significant impacts on the direct
$CP$-violating asymmetry of $\overline{B}^0 \to \pi^{0} \rho^{0}$
decay. This is due to the delicate cancellations among the
competing terms in its decay amplitude, making these sub-leading
contributing terms potentially large. From the numerical results
given in Tables~\ref{tab:cp1} and \ref{tab:cp2}, we can also see
that the higher order penguin contraction contributions to the
direct $CP$-violating asymmetries of $\overline{B}^0 \to \pi^{+}
\rho^{-}$, $B^{-} \to \pi^{-} \omega$, $\overline{B}^0 \to \pi^{-}
\rho^{+}$, and $\overline{B}^0 \to K^{-} \rho^{+}$ decays are also
quite large. In particular, these higher order penguin contraction
contributions can increase the direct $CP$-violating asymmetries
of the former two, while decrease those of the latter two by the
same magnitude.

Although the uncertainties from various input parameters are
reduced to some extent, the renormalization scale dependence of
the direct $CP$-violating asymmetries for some decay modes, such
as $B^{-} \to K^{-} \omega$ and $B^{-} \to K^{-} \rho^{0}$ decays,
are still large. This is due to the fact that the imaginary parts
of the coefficients $\alpha_i$ define by Eq.~(\ref{ais}), which
are crucial for the direct $CP$-violating asymmetries, generally
have a larger scale dependence~\cite{bbns2}.

The direct $CP$-violating asymmetries for some hadronic charmless
$B\to PV$ decays have been measured recently, the data are still
too uncertain to draw any meaningful conclusions from the
comparison with the theoretical predictions, which also suffer
from large uncertainties. With theoretical progresses and the
rapid accumulation of experimental data, the situation will be
improved and large direct $CP$-violating asymmetries in  some
decay channels, for example, the decays $B\to K {\bar K}^\ast$,
will be found in the near future.

\subsection{Detailed analysis of $B\to \pi K^{\ast}$, $K \rho$ decays}
\label{sec3.4}

The $B\to \pi K^{\ast}$ and $B\to K \rho$ decays, like their $PP$
counterparts $B\to \pi K$ decays, are also penguin-dominated decay
modes, and hence sensitive to any new physics contributions. If
the ``$\pi K$'' puzzles, with the improvement of experimental
measurements, still remain unexplained within the SM, there would
be signals of new physics beyond the SM~\cite{Buras:2004th}. Thus,
the $B\to \pi K^{\ast}$ and $B\to K \rho$ decays can be used to
determine whether there are any new physics contributions and,
which one, if exist as hinted by the ``$\pi K$'' puzzles, is more
favored. On the other hand, once the data on these decay modes
becomes more precise, useful information on the weak phase angle
$\gamma$ can also be obtained from these decay
modes~\cite{Sun:2003wn}. So, detailed studies on these decay modes
are worthy.

\begin{figure}[t]
\begin{center}
\scalebox{0.7}{\epsfig{file=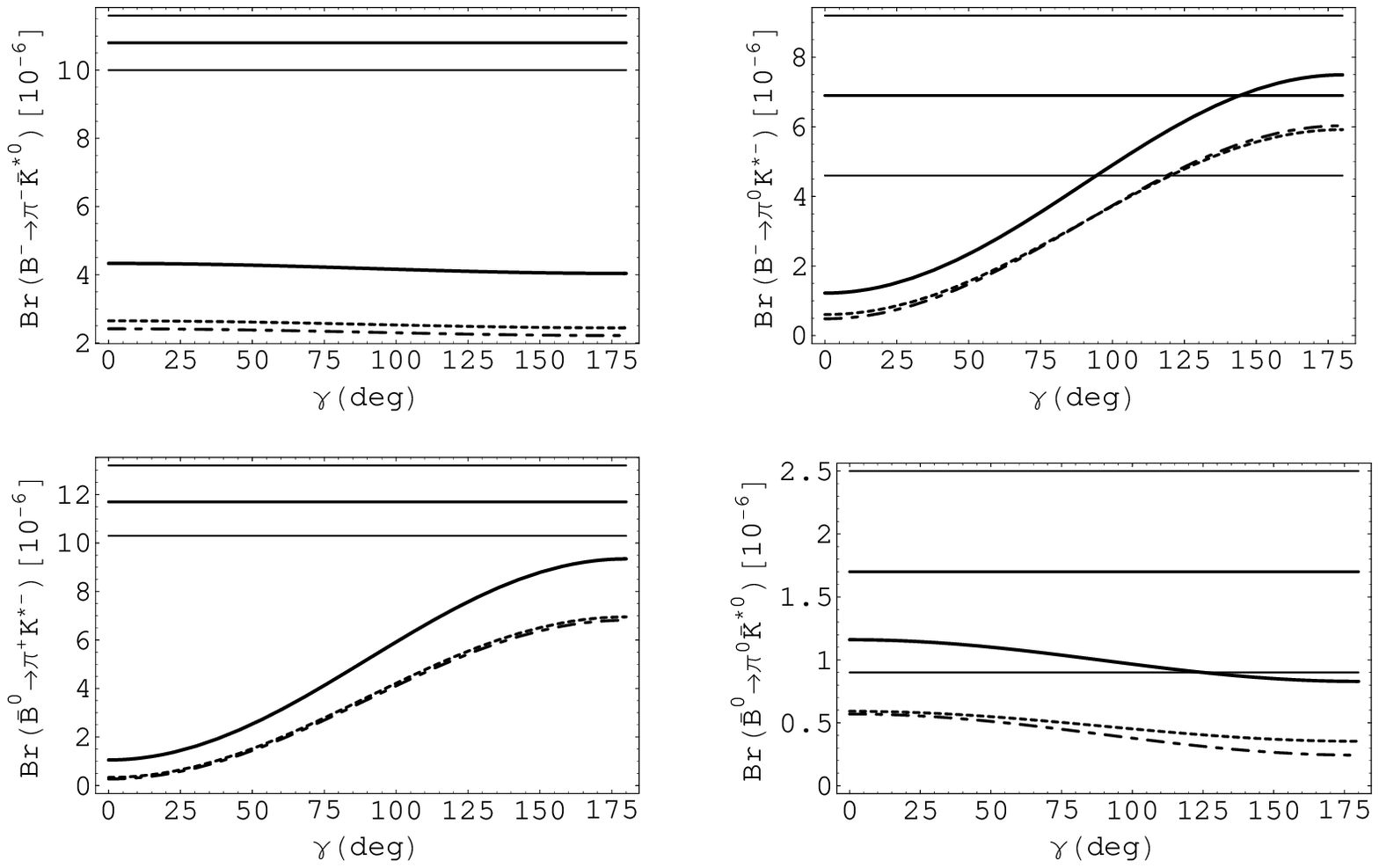}}\\
\scalebox{0.7}{\epsfig{file=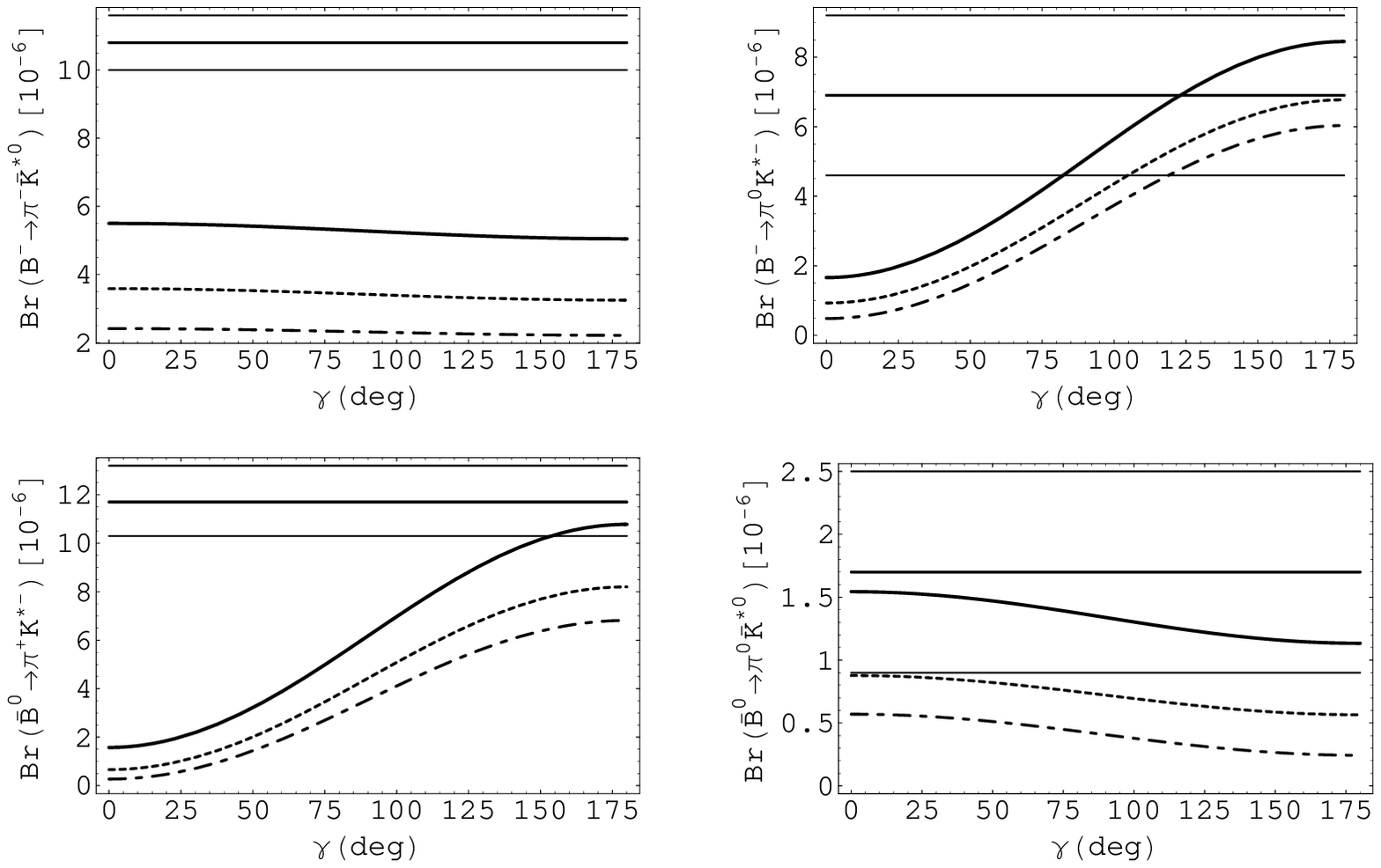}} \caption{\small The
$\gamma$ dependence of the $CP$-averaged branching ratios of $B
\to \pi K^{\ast}$ decays. The upper and the lower four plots
denote the results without and with the annihilation
contributions, respectively. The solid and dashed lines correspond
to the theoretical predictions with and without the higher order
penguin contraction contributions, respectively. The horizontal
solid lines denote the experimental data as given in
Table~\ref{tab:br1}, with the thicker ones being its central
values and the thinner its error bars. The NF results denoted by
the dash-dotted lines are also shown for comparison.}
\label{fig:pik}
\end{center}
\end{figure}

\begin{figure}[t]
\begin{center}
\scalebox{0.7}{\epsfig{file=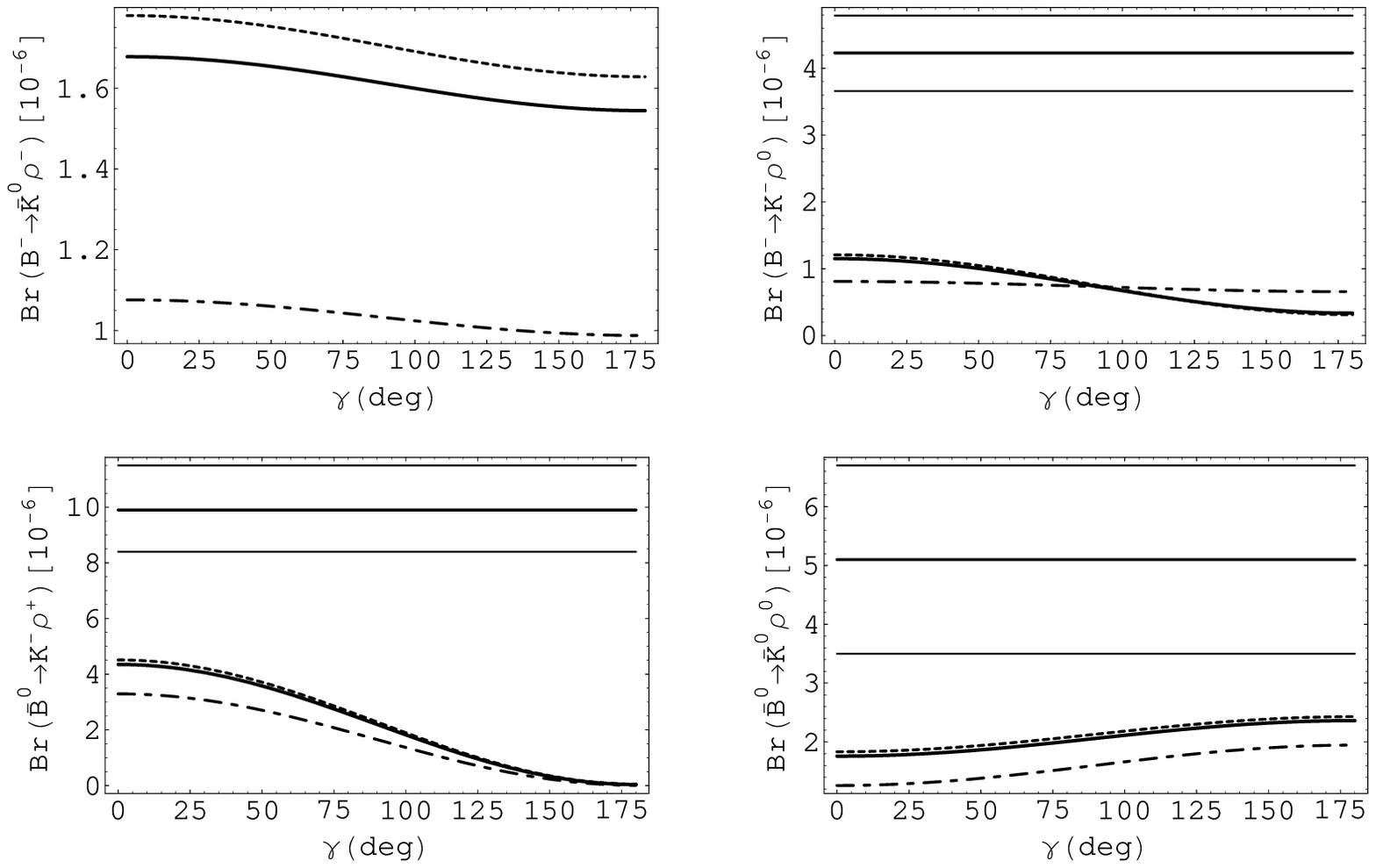}}\\
\scalebox{0.7}{\epsfig{file=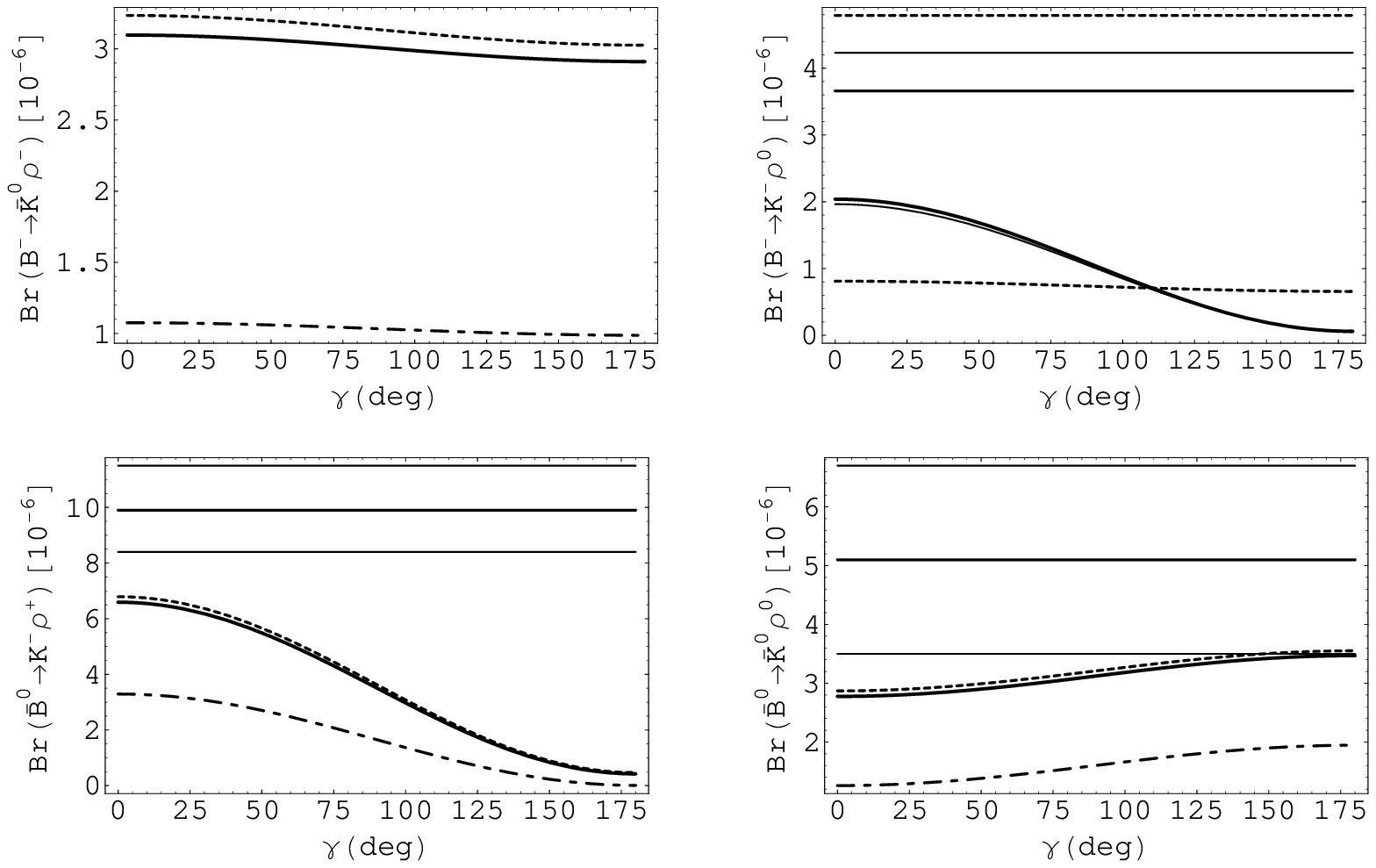}} \caption{The same as
Fig.~\ref{fig:pik} but for $B\to K \rho$ decays.} \label{fig:krho}
\end{center}
\end{figure}

In Figs.~\ref{fig:pik} and \ref{fig:krho}, we show the dependence
of the $CP$-averaged branching ratios of these decay modes on the
weak phase $\gamma$. In these two and the following figures, the
central values of all input parameters except for the CKM angle
$\gamma$ are defaulted and the renormalization scale is fixed at
$\mu=m_b$.


From these two figures, we can see that the experimental data on
these decay modes are generally larger than the theoretical
predictions obtained based on the QCDF approach. Some decay modes,
such as $B^{-}\to \pi^{0} K^{\ast -}$ and $\overline{B}^{0}\to
\pi^{+} K^{\ast -}$ decays, have a strong dependence on the weak
angle $\gamma$. Moreover, both the higher order penguin
contraction and the weak annihilation contributions can give
significant enhancements to the $CP$-averaged branching ratios of
$B\to \pi K^{\ast}$ decays. However, for $B\to \pi K^{\ast}$
decays, only the weak annihilation contributions can provide large
enhancements to the $CP$-averaged branching ratios, and the higher
order penguin contraction contributions play only a minor role.

Since the theoretical uncertainties in the predicted $CP$-averaged
branching ratios can be largely reduced by taking ratios among
them, we shall discuss below certain ratios among the
$CP$-averaged branching fractions of these decay modes, like the
ones defined for $B\to \pi K$ decays~\cite{FM-charge}.

For $B\to \pi K^{\ast}$ decays, we define the following three
ratios~\cite{Sun:2003wn}
\begin{eqnarray}
R(\pi K^{\ast}) &\equiv&
\frac{\tau_{B_u}}{\tau_{B_d}}\,\frac{\bar{{\cal
B}}(\overline{B}^0\to \pi^{+} K^{\ast -})}{\bar{{\cal
B}}(B^{-}\to \pi^{-} \overline{K}^{\ast 0})}\,,\label{R:pik1}\\
R_c(\pi K^{\ast})&\equiv& 2\,\frac{\bar{{\cal B}}(B^{-}\to \pi^{0}
K^{\ast -})}{\bar{{\cal
B}}(B^{-}\to \pi^{-} \overline{K}^{\ast 0})}\,,\label{R:pik2}\\
R_n(\pi K^{\ast})&\equiv& \frac{1}{2}\,\frac{\bar{{\cal
B}}(\overline{B}^0\to \pi^{+} K^{\ast -})}{\bar{{\cal
B}}(\overline{B}^0\to \pi^{0} \overline{K}^{\ast
0})}\,.\label{R:pik3}
\end{eqnarray}
With $\pi$($K^{\ast}$) meson replaced by $\rho$($K$) meson, we can
get another three similar ratios for  $B\to K \rho$ decays. These
ratios should be more appropriate to derive information on the
weak phase angle $\gamma$, as well as the relative strength of
tree and penguin contributions than branching ratios.

\begin{figure}[t]
\begin{center}
\scalebox{0.7}{\epsfig{file=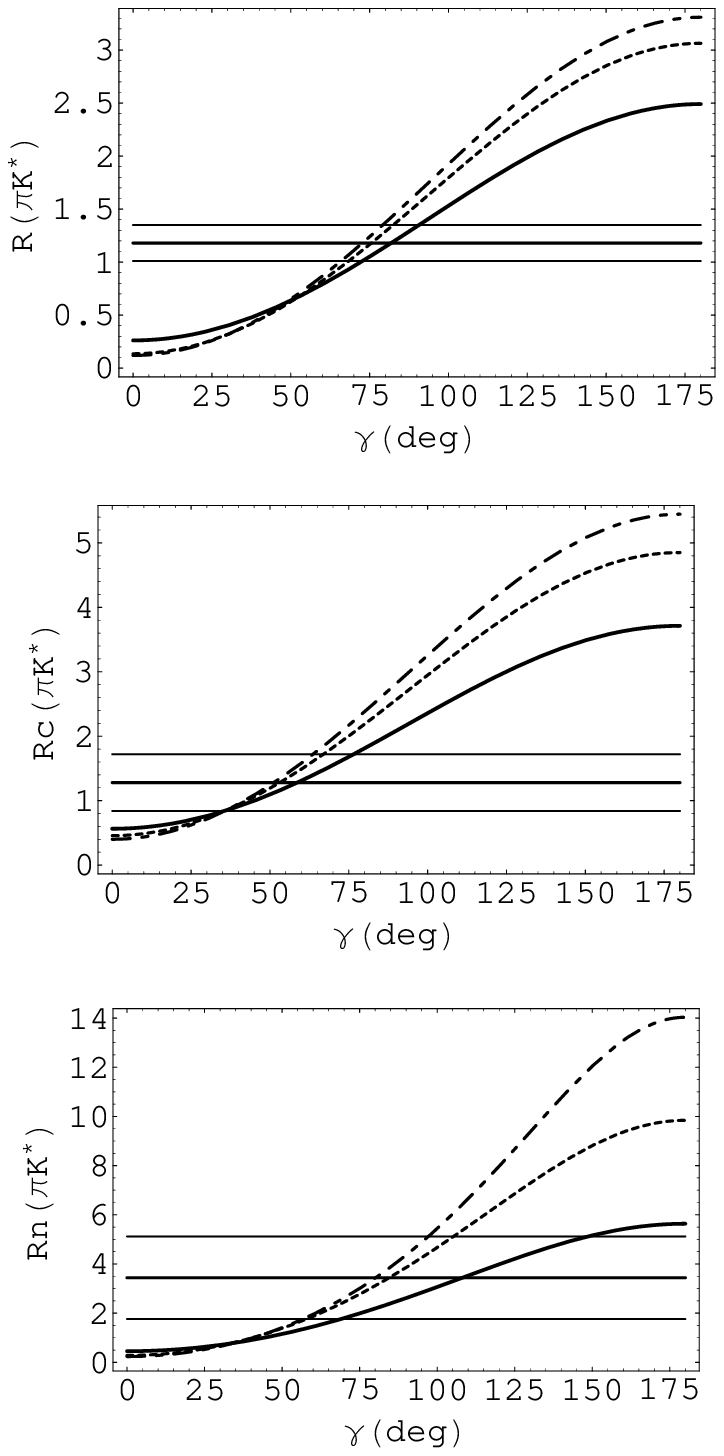}}
\scalebox{0.7}{\epsfig{file=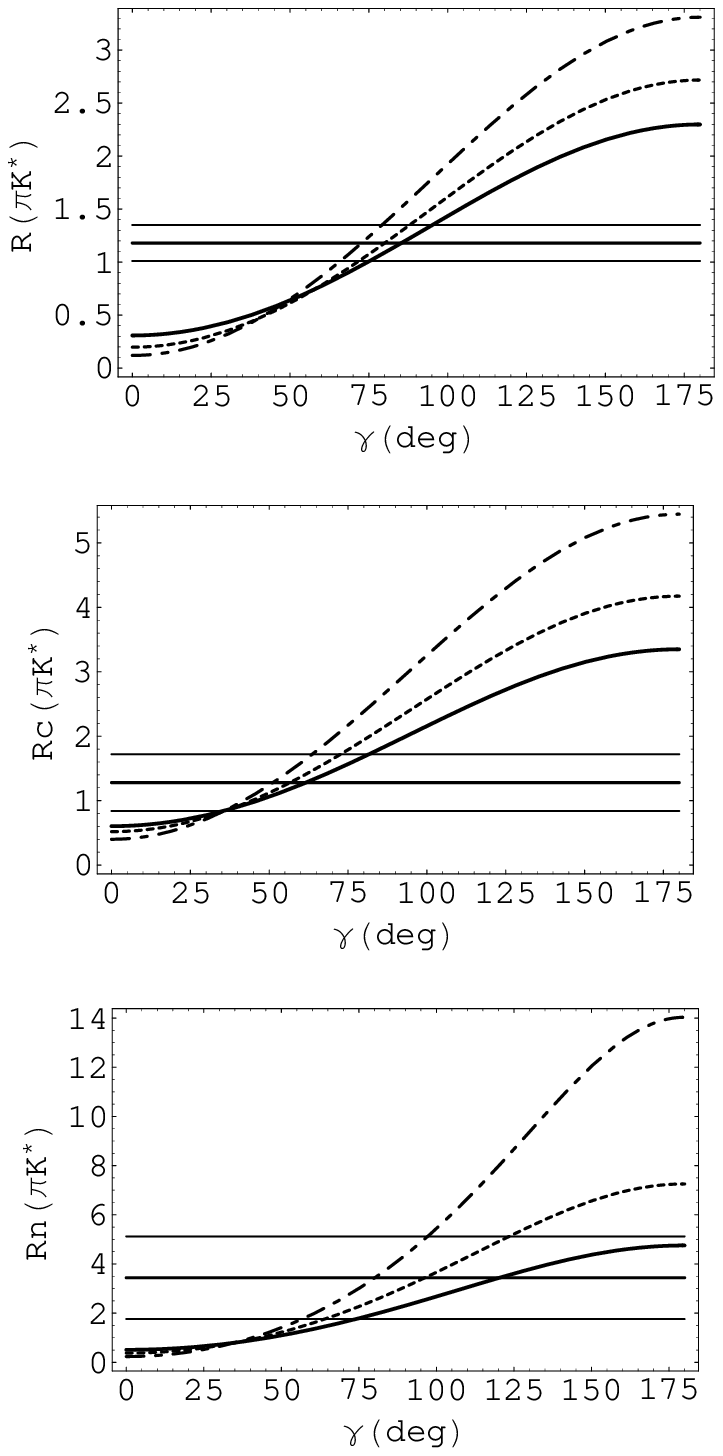}} \caption{ \small Ratios
of the $CP$-averaged branching fractions for $B\to \pi K^{\ast}$
decays defined by Eqs.~(\ref{R:pik1})--(\ref{R:pik3}) as functions
of the weak phase $\gamma$. The left and the right plots denote
the results without and with the annihilation contributions,
respectively. The meaning of the other lines is the same as in
Fig.~\ref{fig:pik}.}\label{fig:Rpik}
\end{center}
\end{figure}

\begin{figure}[t]
\begin{center}
\scalebox{0.7}{\epsfig{file=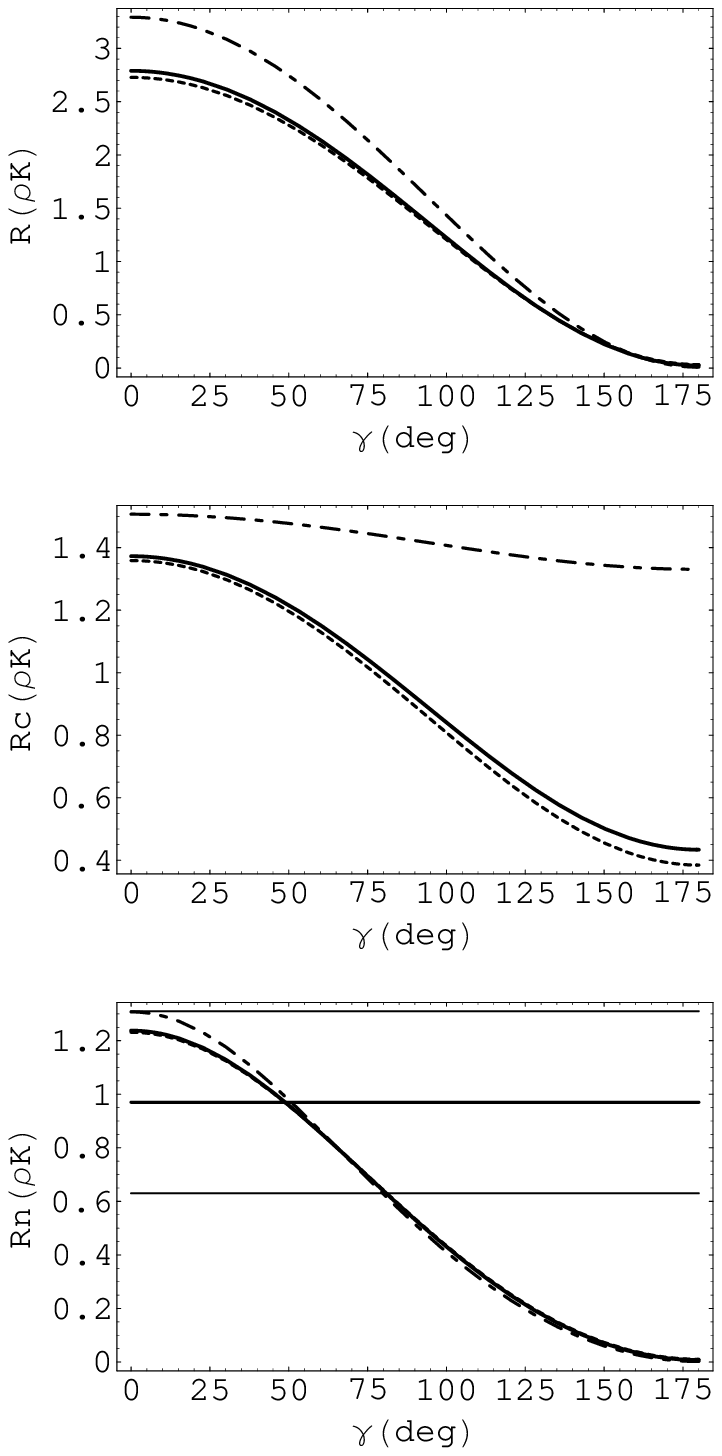}}
\scalebox{0.7}{\epsfig{file=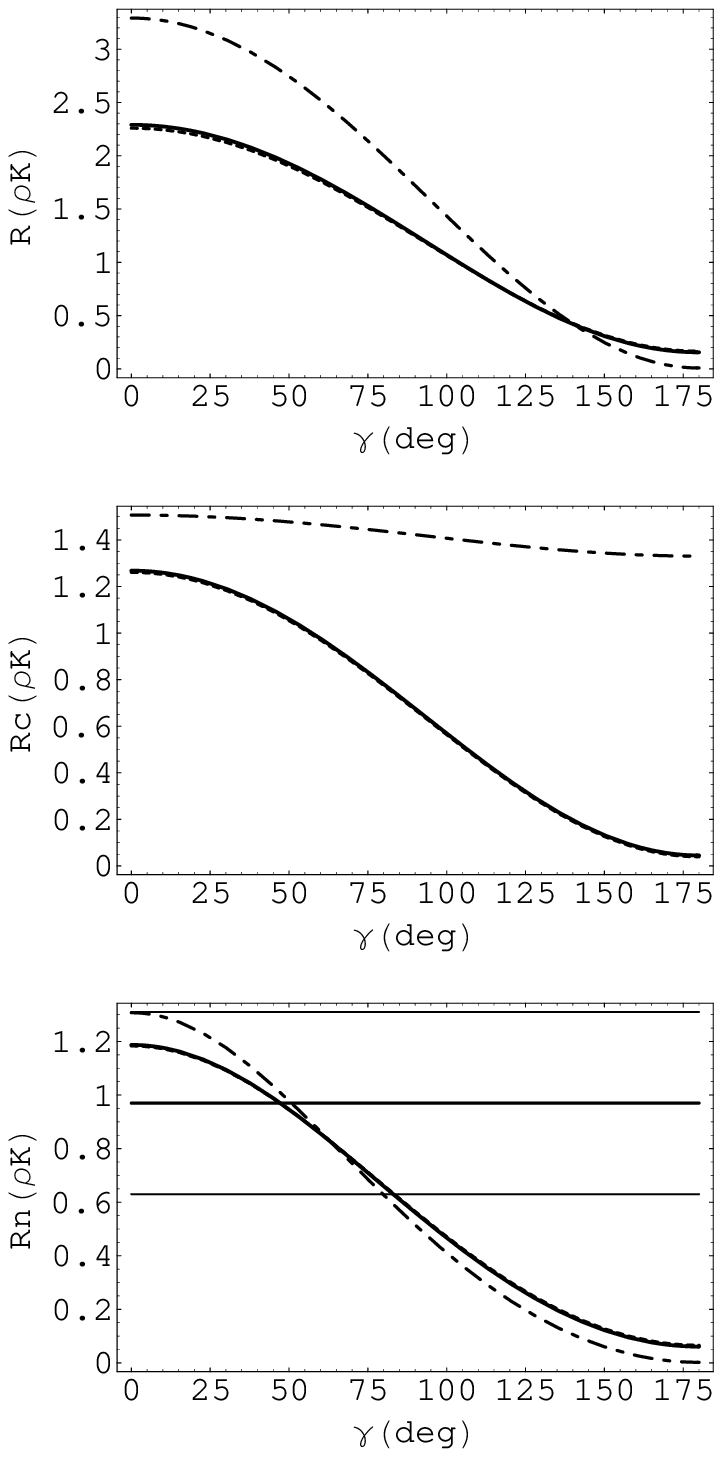}}
\caption{\label{fig:Rkrho} The same as Fig.~\ref{fig:Rpik} but for
$B\to K\rho$ decays}
\end{center}
\end{figure}

Our numerical results and the current experimental data for these
ratios are presented in Table~\ref{tab:Rvalue}. The $\gamma$
dependence of these ratios are displayed in Figs.~\ref{fig:Rpik}
and \ref{fig:Rkrho}.

\begin{table}[t]
\centerline{\parbox{16cm} {\caption{\label{tab:Rvalue} Ratios
among the $CP$-averaged branching fractions of $B\to \pi K^{\ast},
K \rho$ decays. Numbers shown in columns 3 and 4 correspond to the
results obtained without the annihilation contributions, while
those in columns 5 and 6 the ones with the annihilation
contributions. The other captions are the same as in
Table~\ref{tab:br1}.}}}
\begin{center}
\doublerulesep 0.8pt \tabcolsep 0.1in
\begin{tabular}{lcccccc}\hline\hline
 &NF&${\cal O}(\alpha_s)$ & ${\cal O}(\alpha_s+\alpha_s^2)$
    &${\cal O}(\alpha_s)$ & ${\cal O}(\alpha_s+\alpha_s^2)$ & EXP.\\
 \hline
 $R(\pi K^{\ast})$
 &$0.84^{+0.16}_{-0.14}$& $0.80^{+0.18}_{-0.14}$& $0.77^{+0.14}_{-0.11}$
 &$0.76^{+0.15}_{-0.12}$& $0.76^{+0.12}_{-0.11}$& $1.18\pm 0.17$\\
 &$0.75^{+0.11}_{-0.09}$& $0.74^{+0.16}_{-0.11}$& $0.74^{+0.11}_{-0.11}$
 &$0.72^{+0.15}_{-0.10}$& $0.73^{+0.11}_{-0.10}$&\\
 $R_c(\pi K^{\ast})$
 &$1.53^{+0.45}_{-0.31}$& $1.45^{+0.49}_{-0.31}$& $1.28^{+0.29}_{-0.22}$
 &$1.33^{+0.37}_{-0.26}$& $1.22^{+0.25}_{-0.20}$& $1.28\pm 0.44$\\
 &$1.24^{+0.27}_{-0.21}$& $1.32^{+0.45}_{-0.27}$& $1.22^{+0.28}_{-0.20}$
 &$1.24^{+0.32}_{-0.23}$& $1.17^{+0.23}_{-0.19}$&\\
 $R_n(\pi K^{\ast})$
 &$1.87^{+0.94}_{-0.53}$& $1.80^{+1.14}_{-0.53}$& $1.41^{+0.51}_{-0.32}$
 &$1.54^{+0.66}_{-0.42}$& $1.31^{+0.42}_{-0.26}$& $3.44\pm 1.68$\\
 &$1.37^{+0.48}_{-0.29}$& $1.58^{+0.80}_{-0.44}$& $1.33^{+0.50}_{-0.28}$
 &$1.40^{+0.53}_{-0.34}$& $1.25^{+0.43}_{-0.24}$&\\
 $R(\rho K)$
 &$2.55^{+2.45}_{-0.94}$& $2.12^{+1.54}_{-0.67}$& $2.17^{+1.73}_{-0.72}$
 &$1.78^{+0.85}_{-0.41}$& $1.80^{+0.87}_{-0.41}$& $ >0.22$\\
 &$3.20^{+6.80}_{-1.48}$& $2.41^{+2.62}_{-0.85}$& $2.53^{+2.78}_{-0.93}$
 &$1.91^{+1.04}_{-0.50}$& $1.97^{+1.16}_{-0.55}$&\\
 $R_c(\rho K)$
 &$1.47^{+1.96}_{-0.66}$& $1.14^{+0.99}_{-0.41}$& $1.16^{+1.12}_{-0.43}$
 &$0.98^{+0.53}_{-0.26}$& $0.99^{+0.56}_{-0.30}$& $ >0.18$\\
 &$1.52^{+4.78}_{-0.80}$& $1.14^{+1.49}_{-0.49}$& $1.21^{+1.76}_{-0.55}$
 &$0.94^{+0.60}_{-0.31}$& $0.95^{+0.75}_{-0.29}$&\\
 $R_n(\rho K)$
 &$0.88^{+0.44}_{-0.26}$& $0.87^{+0.34}_{-0.23}$& $0.87^{+0.35}_{-0.24}$
 &$0.87^{+0.25}_{-0.20}$& $0.87^{+0.26}_{-0.21}$& $0.97\pm 0.34$\\
 &$0.87^{+0.52}_{-0.26}$& $0.84^{+0.39}_{-0.23}$& $0.85^{+0.40}_{-0.27}$
 &$0.84^{+0.26}_{-0.20}$& $0.84^{+0.26}_{-0.21}$&\\
 \hline\hline
\end{tabular}
\end{center}
\end{table}


From these two figures and the numerical results given in
Table~\ref{tab:Rvalue}, we can see that our theoretical
predictions for most of these ratios are in agreement with the
data, considering the large uncertainties in the experimental
data.

From the explicit expressions of the decay amplitudes for these
decay modes as given, for example, in Ref.~\cite{bbns3}, we can
see that differences between the two ratios $R_c$ and $R_n$ for
both $\pi K^{\ast}$ and $K \rho$ modes arise mainly from the
color-allowed electro-weak penguin coefficient $\alpha_{3,ew}^p$
and the color-suppressed tree coefficient $\alpha_2$, which are
both predicted to be small within the QCDF formalism. So, the two
ratios $R_c$ and $R_n$ are expected to be approximately equal
within the SM. However, due to delicate cancellations among
various competing terms, these ratios are strongly affected by the
sub-leading contributing terms. After including the weak
annihilation contributions, the two ratios $R_c$ and $R_n$ tend to
be approximately equal. The current experimental data, however,
indicate that $R_n(\pi K^{\ast})$ is somewhat larger that $R_c(\pi
K^{\ast})$, but with large errors in the former. Unfortunately,
due to the insufficient data on the branching ratios of the
$K\rho$ modes, direct experimental comparison between $R_c(\rho
K)$ and $R_n(\rho K)$ is not feasible for the time being. Once the
experimental ``$R_c-R_n$'' comparison with the case of $\pi
K^{\ast}$ and $K\rho $ decays are available, we can determine
whether our theoretical predictions based on the QCDF approach are
correct.

It is also noted that the patterns of these quantities remain
nearly unaffected even with these higher order penguin
contributions included, because the higher order penguin
contraction contributions to the decays in the same ratio are
similar in nature, and hence eliminated.

With refined measurements available in the forthcoming years, it
would be very interesting to check whether the theoretical
predictions for these ratios are consistent with the data.
Moreover, studies on these $B\to PV$ modes will help us to
understand the ``$\pi K$'' puzzles~\cite{Buras:2004th}.

\section{Conclusions}\label{sec4}

In this paper, we have reexamined the hadronic charmless $B\to
PV$~(with $P=(\pi,K)$, and $V=(\rho, K^{\ast}, \omega, \phi)$)
decays in the framework of the QCDF.  We  have   taken into
account the penguin contractions of spectator-scattering
amplitudes induced by the $b\to D g^\ast g^\ast$ transitions,
which are of order $\alpha_s^2$. The main conclusions of this
paper are summarized as follows.
\begin{enumerate}

\item For penguin-dominated $B\to PV$ decays, predictions obtained
based on the QCDF approach are generally quite different from the
ones obtained with the NF approximation due to large
``nonfactorizable'' effects on the penguin coefficients. Contrary
to their $PP$ counterparts, the $PV$ modes usually have smaller
penguin coefficients $\alpha_4^p$, rendering the sub-dominant
terms potentially large. For example, the weak annihilation
contributions, though power suppressed by $\Lambda_{\rm QCD}/m_b$,
are very significant in these penguin-dominated decays. The higher
order penguin contraction contributions can interfere
significantly with the next-to-leading order results, and hence
are also important for these penguin-dominated decay modes. In
particular, for decay modes dominated by the coefficient
$\alpha_4^{p}(PV)$, the higher order penguin contraction
contributions can increase the $CP$-averaged branching ratios by
about $40\%\sim 90\%$, while for those dominated by the
coefficient $\alpha_4^{p}(VP)$, their contributions are predicted
to be small and tend to decrease the branching ratios of these
decay modes.

\item For tree-dominated decays and the decays having only the
penguin coefficients $\alpha_3^p$, $\alpha_{3,ew}^p$ or having
only the weak annihilation contributions, the higher order penguin
contraction contributions to the $CP$-averaged branching ratios
are predicted to be quite small.

\item Since the direct $CP$-violating asymmetries are proportional
to the $\sin$ of strong phase, which is usually suppressed by
$\alpha_s$ and/or $\Lambda_{\rm QCD}/m_b$ within the QCDF
formalism, most of the hadronic charmless $B\to PV$ decays are
predicted to have typically small direct $CP$-violating
asymmetries. However, for those decay modes where there are large
interference effects between various contributing terms in the
decay amplitudes, such as $B\to K K^\ast$ decays, large direct
$CP$-violating asymmetries are predicted.

\item Due to large cancellations among the strong phases
associated with the individual Feynman diagram in
Fig.~\ref{penguinfig}, the higher order penguin contraction
contributions to the direct $CP$-violating asymmetries for most
$B\to PV$ decays are predicted to be small, however with a few
exceptions. For example, we find that both the higher order
penguin contraction and the weak annihilation contributions have
significant impacts on the direct $CP$-violating asymmetry of
$\overline{B}^0 \to \pi^{0} \rho^{0}$ decay. In addition, the
higher order penguin contraction contributions to the direct
$CP$-violating asymmetries of $\overline{B}^0 \to \pi^{+}
\rho^{-}$, $B^{-} \to \pi^{-} \omega$, $\overline{B}^0 \to \pi^{-}
\rho^{+}$, and $\overline{B}^0 \to K^{-} \rho^{+}$ decays are also
quite large.

\item With more accurate experimental measurements available in
the forthcoming years, it would be very interesting to check
whether the theoretical predictions for the ratios $R$, $R_c$, and
$R_n$ for both the $\pi K^{\ast}$ and $K\rho$ decay modes are
consistent with the experimental data. In particular, the
experimental $R_c-R_n$ comparison with the case of $\pi K^{\ast}$
and $\rho K$ decays are very crucial for our understandings of the
``$\pi K$'' puzzles.
\end{enumerate}

Although the theoretical results presented here still have large
uncertainties, the penguin contractions of spectator-scattering
amplitudes induced by the $b\to D g^{\ast} g^{\ast}$ transitions,
which are of order $\alpha_s^2$, have been shown to be very
important for two-body hadronic charmless $B\to PV$ decays,
particularly for those penguin-dominated ones. It is very
interesting to note that the 1-loop ($\alpha_s^2$) correction to
the hard spectator scattering in the tree-dominated $B\to \pi\pi$
decays has been performed recently~\cite{Beneke:2005vv}, which
forms another part of the NNLO contribution to the QCD
factorization formula for hadronic $B$-meson decays. Using the
PQCD method, the NLO corrections have also been carried out for
$B\to \pi\pi,\pi K $ and $\rho\rho$ decays very
recently~\cite{lihn}. In addition, much progresses in SCET have
also been made in the past two years~\cite{Bauer:2005kd}. With the
steady progress in experimental measurements at BaBar and Belle,
further systematic studies on these higher order contributions to
the rare hadronic $B$-meson decays are therefore interesting and
deserving.

\section*{Acknowledgments}
We thank Prof. Y.~B. Dai for helpful discussions.  The work is
supported by National Science Foundation under contract
No.~10305003, Henan Provincial Foundation for Prominent Young
Scientists under contract No.~0312001700, and the NCET Program
sponsored by Ministry of Education, China.

\begin{appendix}

\section*{APPENDIX A: ANALYTIC EXPRESSIONS FOR THE
\boldmath $\Delta_i$ FUNCTIONS}

In the NDR scheme, after performing the loop momentum integration,
subtracting the regulator $\epsilon$ using the $\overline{\rm MS}$
scheme, and performing the Feynman parameter integrals, we get the
analytic expressions for the $\Delta_i$ functions appearing in
Eqs.~(\ref{f1}) and (\ref{f2})
\begin{eqnarray}
\Delta i_{5} &=& 2 + \frac{2\,r_1}{r_3}\,\left[G_0(r_1) - G_0(r_1+
r_3)\,\right]\, - \frac{4}{r_3}\,\left[G_{-1}(r_1) -
G_{-1}(r_1 + r_3)\,\right]\,,\\
\Delta i_{6} &=& -2 - \frac{4}{r_3} +
  \frac{2\,r_1\left( 1 +r_3 \right) }{r_3^2}\,G_0(r_1) -
  \frac{2\,\left( r_1 + r_3 + r_1\,r_3 \right) }{r_3^2}\,G_0(r_1 + r_3)\,\nonumber\\
&&\mbox{} +
  \frac{4}{r_3}\left[G_{-1}(r_1) - G_{-1}(r_1+r_3)\,\right]\, -
  \frac{\left( 4 - r_1 \right) \,r_1}{r_3^2}\,T_0(r_1) \,\nonumber\\
&&\mbox{} +
  \frac{\left( 4 - r_1 - r_3 \right) \,\left(  r_1 + r_3 \right)
  }{r_3^2}\,T_0( r_1 + r_3)\,,\\
\Delta i_{23} &=& -2 - \frac{2\,r_1}{r_3}\,\left[G_0(r_1) -
G_0(r_1+ r_3)\,\right]\, + \frac{4}{r_3}\,\left[G_{-1}(r_1) -
G_{-1}(r_1+ r_3)\,\right]\,,\\
\Delta i_{26} &=& -\Delta i_{23}\,,\\
\Delta i_{2} &=&-\frac{22}{9}+ \frac{8}{3}\,\ln \frac{\mu}{m_c} -
  \frac{2\,\left(8+ r_1 \right) }{3\,r_3}\,G_0(r_1) +
  \frac{2\,\left( 8+ r_1 - 2\,r_3 \right) }{3\,r_3}\,G_0(r_1 + r_3)\,\nonumber\\
&&\mbox +
  \frac{4}{r_3}\left[G_{-1}(r_1) - G_{-1}(r_1+r_3)\,\right]\,,\\
\Delta i_{3} &=& \frac{22}{9} + \frac{12}{r_3} +
\frac{4\,r_1}{3\,r_3} - \frac{8}{3}\,\ln \frac{\mu}{m_c} -
  \frac{2\,\left( 7\,r_1 - r_3 - 3\,r_1\,r_3 + 2\,r_1^2 - 2\,r_3^2 \right) }
  {3\,r_3^2}\,G_0(r_1 + r_3)\,\nonumber\\
&& \mbox +
  \frac{2\,r_1\left( 7 + 2\,r_1 -
  3\,r_3 \right) }{3\,r_3^2}\,G_0(r_1) -
  \frac{4\,\left( 2\,r_1 + r_3 \right) }{r_3^2}\,
  \left[G_{-1}(r_1) - G_{-1}(r_1+r_3)\,\right]\,\nonumber\\
&&\mbox +
  \frac{3\,\left( 4 - r_1 \right) \,r_1}{r_3^2}\,T_0(r_1)-
  \frac{3\,\left( 4 - r_1 - r_3 \right) \,\left(  r_1 + r_3 \right)
  }{r_3^2}\,T_0( r_1 + r_3)\,,\\
\Delta i_{8} &=& \frac{32}{9} - \frac{16}{3}\,\ln \frac{\mu}{m_c}-
  \frac{8\left(2 + r_1 \right) }{3\,r_3}\,G_0(r_1)+
  \frac{8\,\left(2 + r_1 +r_3 \right) }{3\,r_3}\,G_0(r_1 + r_3)\,,\\
\Delta i_{12} &=& -\frac{32}{9} + \frac{12}{r_3} +
  \frac{4\,r_1}{3\,r_3} + \frac{16}{3}\,\ln \frac{\mu}{m_c}+
  \frac{2\,r_1\left( 7 + 2\,r_1 + 6\,r_3 \right) }{3\,r_3^2}\,
  G_0(r_1) \,\nonumber\\
&& -
  \frac{2\,\left( 2\,r_1^2 - r_3\,\left( 1 - 4\,r_3 \right)  +
   r_1\,\left( 7 + 6\,r_3 \right)  \right) }{3\,r_3^2} \,G_0(r_1 + r_3)\,\nonumber\\
&& -
  \frac{8\,r_1}{r_3^2}\,\left[G_{-1}(r_1) - G_{-1}(r_1+r_3)\,\right] +
  \frac{3\,\left( 4 - r_1 \right) \,r_1}{r_3^2}\,T_0(r_1) \,\nonumber\\
&& -
  \frac{3\,\left( 4 - r_1 - r_3 \right) \,\left( r_1+ r_3 \right) \,
  }{r_3^2}\,T_0(r_1+r_3) \,\\
\Delta i_{17} &=& \frac{2}{3} + \frac{2\,\left(8+r_1
\right)}{3\,r_3}\,G_{0}(r_1)
 -\frac{2}{3}\,\left(\frac{8+r_1}{r_3}+\frac{4}{r_1+r_3}\right)\,
 G_{0}(r_1 +r_3 ) \nonumber\\
 && -
 \frac{4}{r_3}\,\left[G_{-1}(r_1 )-G_{-1}(r_1 +r_3 ) \,\right]\,,\\
\Delta i_{21} &=& -\frac{2}{3} -\frac{16}{r_3}-\frac{8\,r_1
}{3\,r_3} +\frac{2\,r_1\left( 4\,r_1^2 + 3\,r_3\left( 8 + r_3
\right)  + r_1\left( 20 + 7\,r_3 \right)  \right) }
  {3\,r_3^2\,\left( r_1 + r_3\right) }\,G_{0}(r_1 + r_3 )\,\nonumber \\
&& -
  \frac{2\,r_1\,\left( 20 + 4\,r_1 + 3\,r_3 \right)}{3\,r_3^2}\,G_{0}(r_1 )+
  \frac{4\left( 4\,r_1 + r_3 \right) }{r_3^2}\,
  \left[ G_{-1}(r_1)- G_{-1}(r_1+r_3 )\right]\,\nonumber\\
&& -
  \frac{4\left( 4 - r_1 \right) \,r_1}{r_3^2}\,T_0(r_1) +
  \frac{4\left( 4 - r_1 - r_3 \right) \,\left( r_1+ r_3 \right) \,
  }{r_3^2}\,T_0(r_1+r_3) \,,
\end{eqnarray}
where the notations $r_{1}=k^2/m_c^2$, $r_{2}=p^2/m_c^2$, and
$r_{3}=2\,(k\cdot p)/m_c^2$ have been introduced. With $m_c$
replaced by $m_b$, we can get the results for the b-quark loops.
For light $u, d, s$ quark propagating in the fermion loops, these
$\Delta i$ functions can be evaluated straightforwardly. Here only
the relevant $\Delta i$ functions are given. Explicit expressions
for the remaining ones can be obtained similarly.

The functions $G_i (t)$ and $T_i(t)$ are defined, respectively, by
\begin{eqnarray}
G_{i}(t)&=&\int^1_0 \!dx\, x^i
\ln\left[1-x\,(1-x)\,t-i\delta\,\right]\,,\\
T_{i}(t)&=&\int^1_0 \!dx \frac{x^i}{1-x\,(1-x)\,t-i\delta}\,,
\end{eqnarray}
with the explicit form for $T_0(t)$ given by~\cite{yang:phiXs}
\begin{eqnarray}
T_{0}(t) = \left\{
\begin{array}{cc}
\frac{ 4\,\arctan\sqrt{\frac{t}{4-t}}
}{\sqrt{t\,(4-t)}} ; &0\leq t \leq 4 \\
\frac{2i\,\pi+ 2\ln ( \sqrt{t}-\sqrt{t-4} )-2\ln (
\sqrt{t}+\sqrt{t-4})} {\sqrt{t\,(t-4)}};  &t > 4.
\end{array} \right.\,,
\end{eqnarray}
while the explicit form for $G_{-1,0}(t)$ could be found in
Ref.~\cite{greub}.

\section*{APPENDIX B: INPUT PARAMETERS}

In this appendix, we present the relevant input parameters used in
our numerical calculations as follows.

\indent \textsl{Wilson coefficients.}---The Wilson coefficients
$C_i(\mu)$ have been reliably evaluated to next-to-leading
logarithmic order~\cite{Buchalla:1996vs,Buras:2000}. Their
numerical values in the NDR scheme at the scale
$\mu=m_b$~($\mu_h=\sqrt{\Lambda_h m_b}$ ) are given by
\begin{eqnarray}
 && C_1= 1.080~(1.185),~~
    C_2= -0.180~(-0.367),~~
    C_3=  0.014~(0.028),\nonumber\\
 && C_4= -0.035~(-0.062),~~
    C_5=  0.009~(0.011),~~
    C_6= -0.040~(-0.085),\nonumber\\
 && C_7/\alpha_{\rm e.m.}= -0.009~(-0.029),~~
    C_8/\alpha_{\rm e.m.}=  0.050~(0.107),~~
    C_9/\alpha_{\rm e.m.}= -1.238~(-1.375),\nonumber\\
 && C_{10}/\alpha_{\rm e.m.}= 0.243~(0.451),~~
    C_{7\gamma}^{\rm eff}=-0.302~(-0.365),~~
    C_{8g}^{\rm eff}=-0.145~(-0.169),
\end{eqnarray}
with the input parameters fixed as~\cite{PDG2004}:
${\alpha}_{s}(m_{\rm Z})=0.1187$, ${\alpha}_{\rm e.m.}(m_{\rm
W})=1/129$, $m_{\rm W}=80.425~{\rm GeV}$, $m_{\rm Z}=91.188~{\rm
GeV}$, $\sin^2 \theta_W=0.2312$, $m_{t}=172.7~{\rm
GeV}$~\cite{unknown:2005cc}, $m_{b}=4.65~{\rm GeV}$,
$\Lambda_h=0.5~{\rm GeV}$.

\indent \textsl{The CKM matrix elements.}---Here we use the
Wolfenstein parametrization for the CKM matrix
elements~\cite{Wolfenstein:1983yz}
\begin{eqnarray}
V_{\rm CKM}= \left( \begin{array}{ccc}
1-\frac{\lambda^{2}}{2} &  \lambda &  A\lambda^{3}(\rho-i\eta) \\
-\lambda     & 1-\frac{\lambda^{2}}{2}  & A\lambda^{2}   \\
A\lambda^{3}(1-\rho-i\eta)& -A\lambda^{2}    &   1  \\
\end{array}  \right)\,+{\cal O}(\lambda^4)\,,
\end{eqnarray}
and choose the four Wolfenstein parameters ($A$, $\lambda$,
$\rho$, and $\eta$) as~\cite{Charles:2004jd}
\begin{equation}
A=0.825^{+0.011}_{-0.019},\qquad \lambda=0.22622\pm 0.00100,\qquad
\bar{\rho}=0.207^{+0.036}_{-0.043},\qquad \bar{\eta}=0.340\pm
0.023,
\end{equation}
with $\bar \rho$ and $\bar \eta$ defined by
$\bar{\rho}=\rho\,(1-\frac{\lambda^2}{2})\,,
\bar{\eta}=\eta\,(1-\frac{\lambda^2}{2})$.

\indent \textsl{Masses and lifetimes.}---For the quark mass, there
are two different classes appearing in the QCDF approach. One type
is the pole quark mass which appears in the evaluation of the
penguin loop corrections, and is denoted by $m_q$. In this paper,
we take
\begin{equation}
 m_u=m_d=m_s=0,\qquad m_c=1.46\,{\rm GeV}, \qquad m_b=4.65\,{\rm GeV}.
\end{equation}
The other one is the current quark mass which appears through the
equations of motion and in the factor $r_\chi^M$. This kind of
quark mass is scale dependent. Following Ref.~\cite{bbns3}, we
hold $(\overline{m}_u+\overline{m}_d)(\mu)/\overline{m}_s(\mu)$
fixed, and use $\overline{m}_s(\mu)$ as an input parameter with
the following values
\begin{eqnarray}
2\,\overline{m}_s(\mu)/(\overline{m}_u+\overline{m}_d)(\mu) &=&
24.2\,,\qquad \overline{m}_{s}(2\,{\rm GeV}) = (98\pm20)\,{\rm
MeV}~\cite{Narison:2005ny}\,,\nonumber\\
\overline{m}_{b}(\overline{m}_{b})&=& 4.26\,{\rm
GeV}~\cite{PDG2004}\,,
\end{eqnarray}
where the difference between the $u$ and $d$ quark is not
distinguished.

For the lifetimes and the masses of the $B$ mesons, we choose
\begin{eqnarray}
&&\tau_{B_{u}} = 1.643\,{\rm ps}~\cite{HFAG}\,, \qquad
m_{B_{u}}=5279.0\,{\rm MeV}\,~\cite{PDG2004}\,,\nonumber\\
&&\tau_{B_{d}}=1.527\,{\rm ps}~\cite{HFAG}\,, \qquad
m_{B_{d}}=5279.4\,{\rm MeV}~\cite{PDG2004}\,,
\end{eqnarray}
as our default input values. The masses of the light mesons are
also chosen from Ref.~\cite{PDG2004}.

\indent \textsl{Light-cone distribution amplitudes~(LCDAs) of
mesons.}---The LCDAs of mesons are also basic input parameters in
this approach. In the heavy quark limit, the light-cone projectors
for the $B$, the pseudoscalar, and the vector mesons in the
momentum space can be expressed, respectively,
as~\cite{bbns1,bbns3}
\begin{eqnarray}
 {\cal M}_{\alpha\beta}^B &=& -\frac{i f_B\, m_B}{4}\,\bigg[
 (1+\spur{v} )\,\gamma_5 \,\left\{\Phi_1^B (\xi) + \spur{n_-} \Phi_2^B(\xi)\,
 \right\}\, \biggl]_{\alpha\beta}\,,\label{Bprojector}\\
 M_{\alpha\beta}^P &=& \frac{i f_P}{4} \bigg[\pslash\,\gamma_5\,\Phi_P(x) -
 \mu_P\gamma_5 \frac{\kslash_2\,\kslash_1}{k_1\cdot k_2}\,\Phi_p(x)
   \bigg]_{\alpha\beta}\,,\label{projector1}\\
 (M^V_{\parallel}){\alpha\beta} &=& -\frac{i f_V}{4} \bigg[
   \pslash\,\Phi_V(x) - \frac{m_V \, f_V^{\perp}}{f_V}
   \frac{\kslash_2\,\kslash_1}{k_1\cdot k_2}\,\Phi_v(x)
   \bigg]_{\alpha\beta}\,,\label{projector2}
\end{eqnarray}
where $k_1$ and $k_2$ are the quark and anti-quark momenta of the
meson constituents and defined, respectively, by
\begin{equation}\label{momenta2}
   k_1^\mu = x p^\mu + k_\perp^\mu
    + \frac{\vec k_\perp^2}{2x p\cdot\bar p}\,\bar p^\mu \,, \qquad
   k_2^\mu = (1-x)\, p^\mu - k_\perp^\mu
    + \frac{\vec k_\perp^2}{2\,(1-x)\, p\cdot\bar p}\,\bar
    p^\mu\,.
\end{equation}
It is understood that only after the factor $k_1\cdot k_2$ in the
denominator of Eqs.~(\ref{projector1}) and~(\ref{projector2})
cancelled, can we take the collinear approximation, i.e., the
momentum $k_1$ and $k_2$ can be set to be $x p$ and $(1-x)\, p$,
respectively, with $p$ being the momentum of the meson.
$\Phi_M(x)$ and $\Phi_m(x)$ are the leading twist and twist-3
LCDAs of the meson $M$, respectively. Since the QCDF approach is
based on the heavy quark assumption, to a very good approximation,
we can use the asymptotic forms of the
LCDAs~\cite{Beneke:2000wa,formfactor}~\footnote{It should be noted
~\cite{bbns3,Beneke:2000wa} that, in defining the light-cone
projectors of light mesons, all three-particle
contributions have been neglected. The leading-twist LCDAs are conventionally expanded
in Gegenbauer polynomials
$$
\Phi_M(x,\mu) = 6x(1-x)\,\bigg[ 1 + \sum_{n=1}^\infty
\alpha_n^M(\mu)\,C_n^{(3/2)}(2x-1) \bigg]\,,
$$
where the Gegenbauer moments $\alpha_n^M(\mu)$ are
multiplicatively renormalized. The asymptotic form of the leading
twist distribution amplitude is valid in the limit $\mu \to
\infty$. With three-particle contributions being neglected, the
twist-3 two-particle distribution amplitudes are then determined
completely by the equations of motion.}
\begin{equation}
\Phi_P (x) = \Phi_V (x) =6\,x(1-x),\qquad \Phi_p (x) = 1,\qquad
\Phi_v (x) = 3\,(2\,x-1).
\end{equation}
With respect to the endpoint divergence associated with the
momentum fraction integral over the meson LCDAs, following the
treatment in Refs.~\cite{bbns2,Feldmann:2004mg}, we regulate the
integral with an \textit{ad-hoc} cut-off
\begin{equation}
\int_0^1 \frac{\!dx}{x}\, \to \int_{\Lambda_h/m_B}^1
\frac{\!dx}{x} = \ln \frac{m_B}{\Lambda_h},
 \end{equation}
with $\Lambda_h=0.5\,{\rm GeV}$. The possible complex phase
associated with this integral has been neglected.

As for the $B$ meson wave functions, we need only consider the
first inverse moment of the LCDA $\Phi_1^B (\xi)$ defined
by~\cite{bbns2}
\begin{equation}\label{PhiB1}
  \int_0^1 \frac{d\xi}{\xi}\,\Phi_1^B(\xi)
  \equiv \frac{m_B}{\lambda_B}\,,
\end{equation}
where the hadronic parameter $\lambda_B$ has been introduced to
parameterize this integral. In this paper, we take
$\lambda_B=(460\pm110)\,{\rm MeV}$ as our input
value~\cite{Braun}.

\indent \textsl{Decay constants and transition form factors.}---
The decay constants and the form factors are nonperturbative
parameters and can be determined from experiments and/or
theoretical estimations, such as lattice calculations, QCD sum
rules, etc. For their definitions, we refer the readers to
Refs.~\cite{BSW,Beneke:2000wa,formfactor}. In this paper, we take
the following numerical values for these input parameters
\begin{eqnarray}
 & &f_{\pi}=130.7~{\rm MeV}~\cite{PDG2004},\qquad
    f_{K}=159.8~{\rm MeV}~\cite{PDG2004}, \qquad
    f_{B}=216~{\rm MeV}~\cite{Gray:2005ad}, \nonumber\\
 & &f_{\rho}=205~{\rm MeV}~\cite{BallZwicky},\qquad
    f_{\omega}=195~{\rm MeV}~\cite{BallZwicky},\qquad
    f_{K^{\ast}}=217~{\rm MeV}~\cite{BallZwicky},\nonumber\\
 & &f_{\phi}=231~{\rm MeV}~\cite{BallZwicky}, \qquad
    f_{\rho}^{\perp}(1~{\rm GeV}) =160~{\rm MeV}~\cite{BallZwicky},\qquad
    f_{\omega}^{\perp}(1~{\rm GeV}) =145~{\rm MeV}~\cite{BallZwicky},\nonumber\\
 & &f_{K^{\ast}}^{\perp}(1~{\rm GeV}) =185~{\rm MeV}~\cite{BallZwicky},\qquad
    f_{\phi}^{\perp}(1~{\rm GeV}) =200~{\rm MeV}~\cite{BallZwicky}, \nonumber\\
 & &F^{B\to \pi}_+(0)=(0.258\pm0.031)~\cite{BallZwicky}, \qquad
    F^{B\to {K}}_+(0)=(0.331\pm0.041)~\cite{BallZwicky}, \nonumber\\
 & &A^{B\to \rho}_{0}(0)=(0.303\pm0.028)~\cite{BallZwicky}, \qquad
    A^{B\to K^\star}_{0}(0)=(0.374\pm0.034)~\cite{BallZwicky}, \nonumber\\
 & &A^{B\to \omega}_{0}(0)=(0.281\pm0.030)~\cite{BallZwicky},
 \end{eqnarray}
where the form factors are evaluated at the maximal recoil region.
The dependence of the form factors on the momentum-transfer $q^2$
can be found in Ref.~\cite{BallZwicky}. It should be noted that
the transverse decay constant $f_V^{\perp}$ is scale dependent.

\section*{APPENDIX C:  THE GUAGE INDEPENDENCE}
In this appendix, we present a detail checking of gauge independence of our calculation.  

Firstly, we would check the gauge dependence of Fig.5(b) with the
gluon propagator
\begin{equation}
 D^{\mu\nu}(q^{2})=\frac{1}{q^{2}}\,\left( g^{\mu\nu}-
 \xi\,\frac{ q^{\mu}q^{\nu}}{q^{2}}\right),
\end{equation}
where the factor $-i \delta^{ab}$ has been suppressed, and $\xi$
is the gauge dependent parameter.

Before the light-cone projectors for mesons are sandwiched, the
scattering amplitude of this diagram is read as
\begin{equation}\label{pr1}
 \mathcal{A}\propto
 \biggl[ \bar{v}_{d} (p_{1})\gamma_{\alpha}v_{d}(p_{2})\biggl]
 \left[\bar{u}_{u}(p_{3})\gamma_{\nu} \frac{i}{\spur{ l}}\gamma_{\beta}v_{u}(p_{4})\right]
 \biggl[\bar{u}_{s}(p_{5})\sigma_{\mu\rho}q^{\rho}(1+\gamma_{5})u_{b}(p_{b})\biggl]
 D^{\alpha\beta}(k^{2})  D^{\mu\nu}(q^{2}),
\end{equation}
where the spin indices, color indices, and $SU(3)$ color matrices
have been suppressed.  It is easy to show that
\begin{eqnarray}\label{pr2}
 \biggl[ \bar{v}_{d} (p_{1})\gamma_{\alpha}v_{d}(p_{2})\biggl]
 D^{\alpha\beta}(k^{2})
 &=& \frac{1}{(p_{2}-p_{1})^{2}}\,
 \left\{\biggl[ \bar{v}_{d} (p_{1})\gamma^{\beta}v_{d}(p_{2})\biggr]\,\right. \nonumber\\
 && \left. \qquad \qquad -
 \xi \biggl[ \bar{v}_{d}(p_{1})(\spur{p}_{2}-\spur{p}_{1})v_{d}(p_{2})\biggl]
 \frac{p_{2}^{\beta}-p_{1}^{\beta}}{(p_{2}-p_{1})^{2}}\right\}  \nonumber \\
 &=& \biggl[ \bar{v}_{d} (p_{1})\gamma^{\beta}v_{d}(p_{2})\biggr]\,\frac{1}{(p_{2}-p_{1})^{2}},
\end{eqnarray}
where $k=p_{2}-p_{1}$ is the momentum of the gluon connected to
the spectator  $\bar{d}$ quark, $p_{1}$ and $p_{2}$ are the
momentum of  $\bar{d}$ quark before and after scattering,
respectively. In the last step, we have used  the on-shell
condition $\bar{v}_{d} (p_{1})(\spur{p}_{1}+m_{d}) = (
\spur{p}_{2} +m_{d}) v_{d}(p_{2})=0$.  It is also easy to show
that
\begin{eqnarray}\label{pr3}
 \biggl[\bar{u}_{s}(p_{5})\sigma_{\mu\rho}q^{\rho}(1+\gamma_{5})u_{b}(p_{b})\biggl]
 D^{\mu\nu}(q^{2})
 &=& \frac{1}{q^{2}}\,
 \left\{\biggl[\bar{u}_{s}(p_{5})\sigma^{\nu\rho}q_{\rho}(1+\gamma_{5})u_{b}(p_{b})
 \biggl]\,\right.\nonumber\\
 && \left. \qquad -
 \xi \frac{q^{\nu}}{q^{2}}\,\biggl[\bar{u}_{s}(p_{5})\sigma^{\mu\rho}q_{\mu}
 q_{\rho}(1+\gamma_{5})u_{b}(p_{b})\biggl]\,\right\} \nonumber \\
 &=& \frac{1}{q^{2}}\,\biggl[\bar{u}_{s}(p_{5})\sigma^{\nu\rho}q_{\rho}(1+\gamma_{5})u_{b}(p_{b})\biggl].
\end{eqnarray}
From Eq.~(\ref{pr3}), we can see that the gauge invariance of
$Q_{8g}$ removes the gauge-dependent $\xi$ term.

From Eqs.~(\ref{pr1}), (\ref{pr2}), and (\ref{pr3}), we can see
that the scattering amplitude of Fig.5(b) is independent of the
gauge parameter $\xi$. The above proof could be directly extended
to that of Fig.5(c), Fig.6(b) and (c), since the building blocks
$I^{a}_{\mu}(k)$ and  $\tilde{I}^{a}_{\mu}(k)$ defined by Eq.\ref{Ibuilding}  are also
gauge-invariant, i.e.,
$k^{\mu}I^{a}_{\mu}(k)=k^{\mu}\tilde{I}^{a}_{\mu}(k)=0$.

For Fig.5(a), its amplitude reads
\begin{eqnarray}\label{pr4 }
 \mathcal{A}&\propto&
 \biggl[ \bar{v}_{d}(p_{1})\gamma_{\alpha}v_{d}(p_{2})\biggr]\,
 \biggl[\bar{u}_{u}(p_{3})\gamma_{\nu} v_{u}(p_{4})\biggr]\,
 \biggl[\bar{u}_{s}(p_{5})\sigma_{\mu\rho}q^{\rho}(1+\gamma_{5})u_{b}(p_{b})\biggr]\, \nonumber \\
 &&\times D^{\alpha\alpha'}(k^{2})\, D^{\nu\nu'}(p^{2})\, D^{\mu\mu'}(q^{2})\, V_{\mu'\nu'\alpha'}(q,p,k),
\end{eqnarray}
where $V_{\mu'\nu'\alpha'}(q,p,k)$ is the triple-gluon vertex. One
can observe that the amplitude is gauge independent, because of
\begin{eqnarray}
 &&\xi k^{\alpha} \biggl[ \bar{v}_{d}(p_{1})\gamma_{\alpha}v_{d}(p_{2})\biggr] =0, ~~ (k=p_{2}-p_{1}),
 \label{pr5} \\
 &&\xi p^{\nu} \biggl[\bar{u}_{u}(p_{3})\gamma_{\nu} v_{u}(p_{4})\biggr]=0,  ~~(p=p_{3}+p_{4}),
 \label{pr6} \\
 &&\xi q^{\mu}\biggl[\bar{u}_{s}(p_{5})\sigma_{\mu\rho}q^{\rho}(1+\gamma_{5})u_{b}(p_{b})\biggr]=0.
\end{eqnarray}
Similarly, we can find that the amplitude of Fig.6(a) is also
gauge independent. Using Eqs.~(\ref{pr5}) and (\ref{pr6}), one can
find that the amplitudes of Fig.6(d) and (e) are gauge
independent.

In summary, we have shown that the amplitudes of the Feynman
diagrams in Figs.5 and 6 are gauge independent. The gauge independence of this subset Feynman diagrams 
is guaranteed  by the on-shell external quarks and the gauge-invariance of $\mathcal{O}_{8g}$ and  4-quarks operation insertions $I^{a}_{\mu}(k)$ and  $\tilde{I}^{a}_{\mu}(k)$.    There are many
${\cal O}(\alpha^{2}_{s})$ Feynman diagrams belonging to the group
in Fig.4 but  not shown there.  For example, the processes $b\to s
g^{*} $ followed by $g^{*}\to\bar{q}_{i}q_{i}\to g^{*}\to
\bar{u}{u}$ or a gluon loop, which are gauge dependent separately.
To keep gauge-independent, one must calculate the full set of
Feynman diagrams in the category of Fig.4, which are not
calculated in this paper.

\end{appendix}

\end{document}